\documentclass[superscriptaddress,nofootinbib]{revtex4}

\usepackage{amssymb}
\usepackage{epsfig}

\begin{document}

\title{Probabilistic Clustering of Sequences:\\ Inferring new bacterial
regulons by comparative genomics.}
\author{Erik van Nimwegen}
\email{erik@golem.rockefeller.edu}
\homepage{http://physics.rockefeller.edu/~erik}
\affiliation{Center for Studies in Physics and Biology, The Rockefeller University, 1230 York Avenue, New York 10021 NY}
\author{Mihaela Zavolan}
\affiliation{Laboratory of Computational Genomics, The Rockefeller University, 1230 York Avenue, New York 10021 NY}
\author{Nikolaus Rajewsky}
\affiliation{Center for Studies in Physics and Biology, The Rockefeller University, 1230 York Avenue, New York 10021 NY}
\author{Eric D. Siggia}
\affiliation{Center for Studies in Physics and Biology, The Rockefeller University, 1230 York Avenue, New York 10021 NY}

\begin{abstract}
Genome wide comparisons between enteric bacteria yield large sets of
conserved putative regulatory sites on a gene by gene basis that need
to be clustered into regulons. Using the assumption that regulatory
sites can be represented as samples from weight matrices we derive a
unique probability distribution for assignments of sites into
clusters. Our algorithm, 'PROCSE' (probabilistic clustering of
sequences), uses Monte-Carlo sampling of this distribution to
partition and align thousands of short DNA sequences into
clusters. The algorithm internally determines the number of clusters
from the data, and assigns significance to the resulting clusters. We
place theoretical limits on the ability of any algorithm to correctly
cluster sequences drawn from weight matrices (WMs) when these WMs are
unknown. Our analysis suggests that the set of all putative sites for
a single genome (e.g. E. coli) is largely inadequate for
clustering. When sites from different genomes are combined and all the
homologous sites from the various species are used as a block,
clustering becomes feasible. We predict 50-100 new regulons as well as
many new members of existing regulons, potentially doubling the number
of known regulatory sites in E. coli.
\end{abstract}

\maketitle

\tableofcontents

\section{Introduction}

New microbial genomes are sequenced almost daily, and the first step
in their annotation is the elucidation of their protein coding
regions. The noncoding regions of the genome can provide clues about
gene regulation since they contain various regulatory elements. These
elements are generally much smaller and more variable than typical
coding regions, and thus harder to identify. Computational methods are
needed, since even for E. coli, there are only 60-80 genes for which
binding sites and regulated genes are known \cite{church,regulonDB},
whereas protein sequence homology suggests there are $\sim$300 DNA
binding proteins \cite{salgado}. Binding sites have been identified
experimentally in only 300 of the 2200 regulatory regions of E. coli
\cite{regulonDB}. For important pathogens such as V. cholera, Y. pestis,
or M. tuberculosis very little is known about gene regulation from
direct experimentation.

Computational strategies for the discovery of regulatory sites began
with algorithms \cite{consensus,Gibbssampler,meme} which identified
sets of similar sequences in the regulatory regions of functionally
related groups of genes. More recently, algorithms were proposed to
identify repetitive patterns within an entire genome
\cite{bussemaker}. Here we develop methods for partitioning a large
set of putative regulatory sites into clusters based on sequence
similarity, with the goal of identifying regulons. That is, we aim to
partition the set of sites such that each cluster corresponds to those
targeted by the same transcription factor (TF).

Many authors have noted the potential of interspecies comparisons to
elucidate regulatory motifs, e.g. \cite{hardison}. Generally, a group
of functionally related genes in bacteria is pooled to extract common
sites within the regulatory regions of these genes,
e.g. \cite{gelfand,mcguire}. More recent studies
\cite{McCueEtAl2001,RajewskyEtAl2001} have shown that when upstream
regions of orthologous genes from several suitably related species are
compared at once, there is sufficient signal for regulatory sites to
be inferred on a gene by gene basis, yielding thousands of potentially
new sites. These sites form the data sets on which our algorithm
operates.

Previous algorithms that fit WMs can not process genome scale data
representing sites from hundreds of TFs simultaneously. Other schemes
\cite{bussemaker}, not based on WM representations of regulatory
sites, are not well suited for processing sites that were inferred
from interspecies comparison. Our algorithm partitions the entire set
of sites at once, infers the number of clusters internally, and
assigns probabilities to all partitions of sequences into
clusters. Within this framework, we also derive theoretical limits on
the {\em clusterability} of sets of regulatory sites. 

A set of sites, sampled from a set of {\em unknown} WMs, is said to be
clusterable if it is possible to infer which sites where sampled from
the same WM. If the WMs from which the sites were sampled are {\em
known}, we have the much simpler classification problem: determining
which sites were sampled from which WM. It is important to realize
that the cell is performing a classification task since it {\em knows}
the WMs of the TFs, i.e. the chemistry of the DNA-protein interaction
automatically assigns a binding-energy to each site just as a WM
assigns a score to each site. However, since we cannot infer binding
specificities from a TF's protein sequence, we face the much harder
clustering task. Our theoretical arguments and the available data for
E. coli in fact suggest that the set of all regulatory sites in the
E. coli genome is unclusterable by itself. However, we also show how
this problem can be circumvented by taking into account information
from interspecies comparison.

\section{Model}

Protein binding sites in bacterial genomes are commonly described by a
WM, $w^i_{\alpha}$, which gives the probabilities of
finding base $\alpha$ at position $i$ of the binding
site \cite{von_hippel}. The probabilities in different columns $i$ are
assumed independent, which accords well with existing compilations
\cite{church}.  Motif-finding algorithms
\cite{consensus,Gibbssampler,meme} score the quality of an alignment
of putative binding sites by the information score $I$ of its
(estimated) WM:
\begin{equation}
\label{information_score}
I = \sum_{i,\alpha} w^i_{\alpha} \log(w^i_{\alpha}/b_{\alpha}),
\end{equation}
where $b_{\alpha}$ is the background frequency of base $\alpha$, and
the $w^i_{\alpha}$ are the WM probabilities, estimated from the
sequences in the alignment. The rationale for this scoring function is
that the probability of an $n$-sequence alignment with frequencies
$w^i_{\alpha}$ arising by chance from $n$ independent samples of the
background distribution of bases $b_{\alpha}$ is given by $P \approx e^{-n I}$. 

Instead of distinguishing sequence-motifs for a single TF against a
background distribution, our task is to cluster a set of binding sites
of an unknown number of different TFs, i.e. a set of sequences sampled
from an unknown number of unspecified WMs. To this end, we consider
all ways of {\em partitioning} our data set into clusters and assign a
probability to each partition. 
\begin{figure}[htbp]
\centerline{\epsfig{file=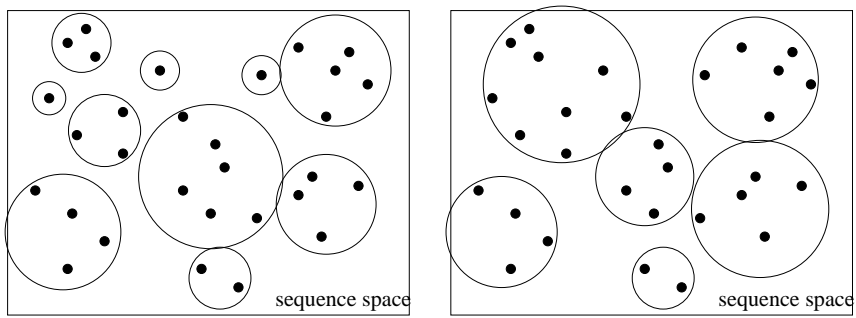,width=4.5in}}
\caption{Two ways of partitioning the same set of sequences into
clusters. The rectangle schematically represents the space of all
possible DNA sequences of some particular length $l$. The dots denote
the sequences in the data set, and the circles indicate which
sequences are partitioned together into clusters.}
\label{partition_examples_fig}
\end{figure}
Figure \ref{partition_examples_fig} depicts, schematically, two ways
of partitioning a set of sequences into clusters. We will assign
probabilities to all such partitions. The probability of a partition
is the product of the probabilities, for each cluster, that all
sequences within the cluster arose from a common WM.

To calculate these probabilities, consider first the conditional
probability $P(S|w)$ that a set of $n$ length-$l$ sequences $S$ was
drawn from a given WM $w$:
\begin{equation}
\label{single_set_prob}
p(S|w) = \prod_{s \in S} \prod_{i=1}^l w^i_{s_i},
\end{equation}
where $s_i$ is the letter at position $i$ in sequence $s$.  The
probability $P(S)$ that all sequences in $S$ came from {\em some} $w$
can be obtained by integrating over all allowed $w$; viz. over the
simplex $\sum_{\alpha} w^i_{\alpha} = 1$ for each position
$i$. Lacking any knowledge regarding $w$ we use a uniform prior over
the simplex. We obtain
\begin{equation}
\label{p_single_cluster}
P(S) = \int  P(S|w) dw = {n+3 \choose 3}^{-l} \prod_{i=1}^l \frac{\prod_{\alpha}
n^i_{\alpha}!}{n!},
\end{equation}
where $n^i_{\alpha}$ is the number of occurrences of base $\alpha$ in
column $i$. The last factor in Eq. \ref{p_single_cluster} is just the
inverse of the multinomial factor that counts the number of ways of
constructing a specific vector $(n_a,n_c,n_g,n_t)$ from $n$ bases,
which bears an obvious relation to Eq. \ref{information_score}. High
probabilities are thus given to vectors which can be realized in the
least number of ways.  The factor ${n+3 \choose 3}$, counts the number
of distinct vectors $(n_a,n_c,n_g,n_t)$ that can be obtained from $n$
samples.

We can now define for any partition $C$ of a data set of sequences $D$
into clusters $S_c$ the likelihood $P(D|C)$ that all sequences in each
$S_c$ were drawn from a single WM: $P(D|C) = \prod_c P(S_c)$, with
$P(S_c)$ given by Eq. \ref{p_single_cluster}. Then the posterior
probability $P(C|D)$ for partition $C$ given the data $D$ is
\begin{equation}
\label{probtot}
P(C|D) = \frac{P(D|C) \pi(C)}{\sum_{C'} P(D|C') \pi(C')},
\end{equation}
where $\pi(C)$ is the prior distribution over partitions, which we
will assume to be uniform.

Consider the simplest example of a data set of only two sequences with
matching bases in $b$ of their $l$ positions. We have $P = 2^b
(1/20)^l$ for the probability that the sequences came from the same
WM, while $P = (1/16)^l$ for the probability that they came from
different WMs. $P(C|D)$ will thus prefer to either cluster or separate
the two sequences, depending on $b$. In general, the probability
distribution $P(C|D)$ will prefer partitions in which similar
sequences are co-clustered. The state space of all partitions (the
number of which grows nearly as rapidly as $n!$ \cite{DeBruijnbook})
acts as an 'entropy' which opposes (stable) clustering of similar
sequences.

The probability distribution Eq. \ref{probtot} allows us to calculate
any statistic of interest by summing over the appropriate partitions
$C$. For instance, to calculate the probability that the data set
separates into $n$ clusters, one sums $P(C|D)$ over all partitions
that contain $n$ clusters. Analogously, we can calculate the
probability that any particular subset of sequences forms a cluster by
summing $P(C|D)$ over all partitions in which this occurs. Our
clustering framework is novel in that it allows for direct
calculations of these quantities. In the implementation section below
we describe how we sample $P(C|D)$ and identify significant clusters
by finding subsets of sequences that consistently cluster.

Generalizations to data arising from WMs of different lengths and
sequences that are not consistently aligned are straightforward and
considered below. It is also trivial to incorporate prior information
on the number of clusters (e.g. that it should equal the number of
TFs).

\section{Classifiability vs. Clusterability}

Correct regulation of gene expression requires that TFs should bind
preferentially to their own sites. Associating TFs with WMs, $P(s|w)$
is commonly taken to be the probability that $w$ binds to $s$. Correct
regulation thus implies that for a sample $s$ from $w$, we have that
$P(s|w) > P(s|w')$ for all other TFs $w' \neq w$, which we call a {\em
classification} task. Formally, we are given a set of WMs and a set of
sequences sampled from them, and assign each sequence $s$ to the WM
from the set that maximizes $P(s|w)$. We define the data to be
classifiable when, in at least half of the cases, the WM $w$ which
maximizes $P(s|w)$ is the WM from which $s$ was sampled. As mentioned
in the introduction, classification is much simpler than clustering a
set of sites in absence of knowledge of the set of WMs from which they were
sampled.

To quantify clusterability, assume we are clustering $n G$ sequences,
that were obtained by sampling $n$ times from each of $G$ different
WMs. For each of these WMs we can calculate the probability that $m$
of its $n$ samples co-cluster by summing the probabilities $P(C|D)$
over all partitions $C$ in which $m$, and no more than $m$, samples of
$w$ occur together in any of the clusters. We will define the set to
be `clusterable' if, for more than half of the $G$ WMs, the average of
$m$, $\langle m \rangle > n/2$.

\begin{figure}[htbp]
\centerline{\epsfig{file=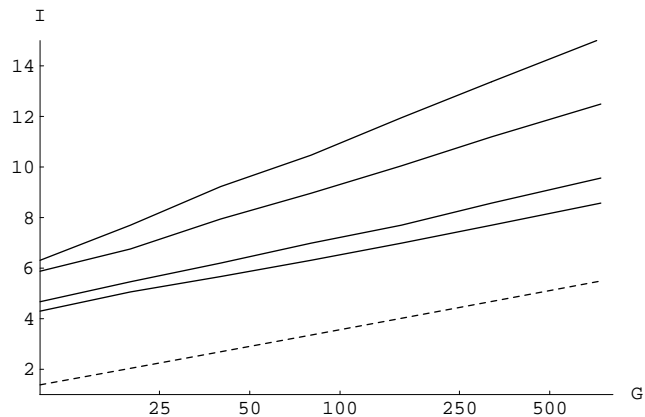,height=5.5cm}}
\caption{The critical information score $I$ for clusterability (solid
lines) or classifiability (dashed line) as a function of the number of
clusters $G$ (shown on a log scale). The solid lines correspond, from
top to bottom, to sets of $n=3$, $5$, $10$, and $15$ samples per
cluster. The WM length is $l=27$.}
\label{clusterability_fig}
\end{figure}

We have performed analytical and numerical calculations that identify
under what conditions a data set is classifiable and clusterable. This
theory is beyond the scope of this paper and will be reported
elsewhere. The results are summarized in
Fig. \ref{clusterability_fig}.
Given the information score $I$ (Eq. \ref{information_score}) of a WM,
the fraction of the space of $4^l$ sequences filled by the binding
sites for this WM is $e^{-I}$. One can thus regard $I$ as a measure of
the specificity of a WM. Fig. \ref{clusterability_fig} shows the
minimal WM specificity necessary to cluster (solid lines) or classify
(dashed line) as a function of the number of WMs $G$ and the number of
samples $n$ per WM. Fig. \ref{clusterability_fig} shows that $\exp(-I)
\propto 1/G$ for classification and $\exp(-I) \approx 1/G^2$ for
clustering a set of $n=3$ binding sites, with fractional exponents in
between these extremes. Thus, all $G$ WMs together consume a fixed
fraction of sequence space at the classification threshold
(independent of $G$), while it decreases as a function of $G$ at the
clusterability threshold. Moreover, there is a significant gap between
the requirements for classification vs. clustering, even for large
numbers of samples. Thus, clustering is impossible for data sets close
to the classification threshold. Results presented below suggest that
the collection of E. coli binding sites may well be in this
unclusterable regime, where few regulons can be correctly inferred.

However, comparative genomic information can salvage this
situation. The putative binding sites of our data sets were extracted
by finding conserved sequences upstream of orthologous genes of
different bacteria (see below). Such conserved sequence sets are
likely to contain binding sites for the {\em same} TF, and should be
clustered together. Therefore, we can significantly reduce the size of
the state space by pre-clustering these conserved sites into so called
mini-WMs, and instead of clustering single sequences, we will be
clustering these mini-WMs using the same probabilities
Eq. \ref{p_single_cluster}. This improves clusterability dramatically.

\section{Implementation}

\begin{figure}[htbp]
\centerline{\epsfig{file=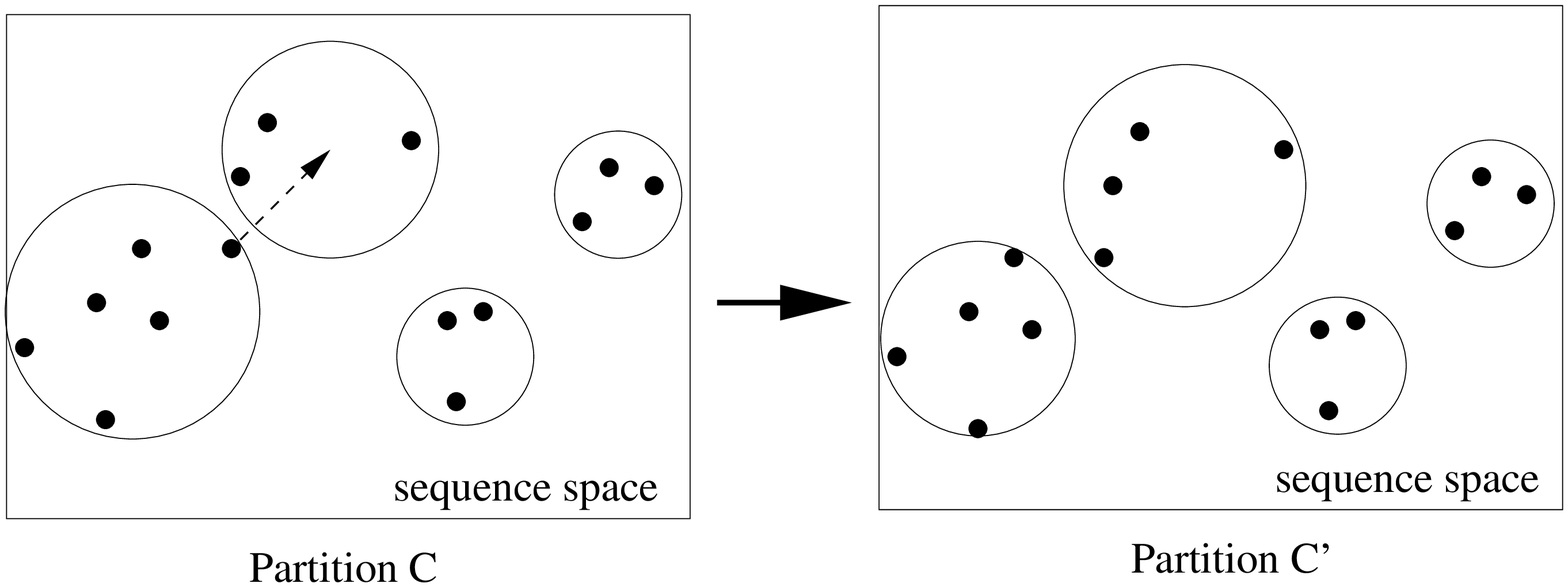,width=4.5in}}
\caption{Monte-Carlo sampling of partitions: example of a move from
partition $C$ to partition $C'$. The dots are sequences and the
circles delineate the clusters.}
\label{move_example}
\end{figure}

We have implemented a Monte-Carlo random walk to sample the
distribution $P(C|D)$. At every 'time step' we choose a mini-WM at
random and consider reassigning it to a randomly chosen cluster (or
empty box). These moves are accepted according to the
Metropolis-Hastings scheme \cite{Metropolis53}: moves that increase
the probability $P(C|D)$ are always accepted and moves that lower
$P(C|D)$ are accepted with probability
$P(C'|D)/P(C|D)$. Fig. \ref{move_example} shows an example of a move
from a partition $C$ to a partition $C'$.
This sampling scheme thus generates 'dynamic' clusters whose
membership fluctuates over time. Clusters may evaporate altogether and
new clusters may form when a pair of mini-WMs is moved together. We
wish to identify 'significant' clusters by finding sets of mini-WMs
that are persistently grouped together during the Monte-Carlo
sampling. Ideally, we would find a set of clusters, each with stable
'core' members that are present at all times, while the remaining
mini-WMs move about between different clusters. Reality is
unfortunately more complicated. One finds clusters that are constantly
drifting, such that their membership is uncorrelated on long time
scales. Other clusters, with stable membership, may evaporate and
reform many times. While we can easily sample $P(C|D)$ to obtain
significance measures for any given 'candidate cluster', the
rich dynamics of drifting, fusing, and evaporating clusters makes it
nontrivial to identify good candidate clusters.

\begin{figure}[htbp]
\centerline{\epsfig{file=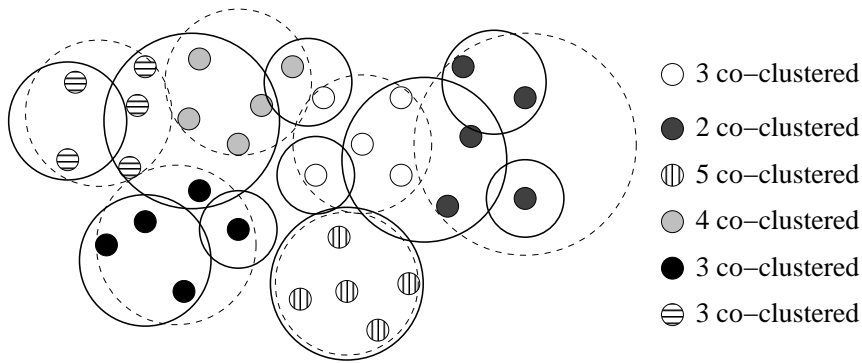,width=4.5in}}
\caption{The ML partition obtained by annealing is indicated by the
thin dashed circles and the fill patterns of the dots. The thick lines
show an alternative partition that may arise during sampling. The
number of co-clustering members in this partition are shown on the
right for each of the ML clusters.}
\label{sig_test_explain}
\end{figure}

We have experimented with a number of schemes for identifying
candidate clusters (see appendix \ref{extract_sig_clusters}). One
approach is to search for the {\em maximum likelihood} (ML) partition
that maximizes Eq. \ref{probtot}. This can be done by simulated
annealing: we raise $P(D|C)$ to the power $\beta$, increasing $\beta$
over time (in practice $\beta = 3$ is large enough). The ML partition
gives us a set of candidate clusters. The significance of the ML
clusters are then tested by sampling
$P(C|D)$. Fig. \ref{sig_test_explain} illustrates this procedure.
For each partition encountered during the sampling, we define the
number of co-clustering members of a ML cluster as the maximum number
of mini-WMs from the ML cluster that co-occur in a single cluster (see
Fig. \ref{sig_test_explain}). In this way we measure, for each ML
cluster, the probabilities $p(k)$ that $k$ of its members
co-cluster\footnote{The mean size of the cluster is thus $\sum_k k
p(k)$.}. Finally, we calculate the minimal length interval $[k_{\rm
min},k_{\rm max}]$ for which $\sum_{k=k_{\rm min}}^{k_{\rm max}} p(k)
> 0.95$. All clusters for which $k_{\rm min} \geq 2$ are deemed
significant.

This method is computationally prohibitive for large data sets
(because we cannot run long enough to converge {\em all} cluster
statistics). For larger data sets we measure, using several
Monte-Carlo random walks, the probability that each pair of mini-WMs
co-clusters\footnote{Note that these pair statistics {\em can not} be
calculated in terms of the sequences in the pair of mini-WMs
themselves. They depend on the full data-set.}. We then construct a
graph in which nodes correspond to mini-WMs, and edges between
mini-WMs $i$ and $j$ exist if and only if their co-clustering
probability $p_{ij} > 1/2$. Candidate clusters are now given by the
connected components of this graph. The pairwise statistics are then
further processed to obtain probabilistic cluster membership. This
yields for each mini-WM $i$, the probabilities $p^i_j$ that mini-WM
$i$ belongs to cluster $j$ (see appendix \ref{extract_sig_clusters}). We also
calculate, for each cluster, the probability distribution $p(k)$ of
$k$ of its members co-clustering. Cluster significance is judged from
$p(k)$ as described above. Fortunately, there is substantial agreement
on the significant clusters among these ways of extracting significant
clusters from $P(C|D)$.

After we have inferred the clusters and their members, we can estimate
a WM for each cluster. We then classify all mini-WMs in the full data
set in terms of these cluster WMs. Finally, we search for additional
matching motifs to the cluster WMs in all the regulatory regions of
the E. coli genome. Details for all these procedures are described in
appendix \ref{extract_sig_clusters}.

\section{Data Sets}

\begin{figure}[htbp]
\centerline{\epsfig{file=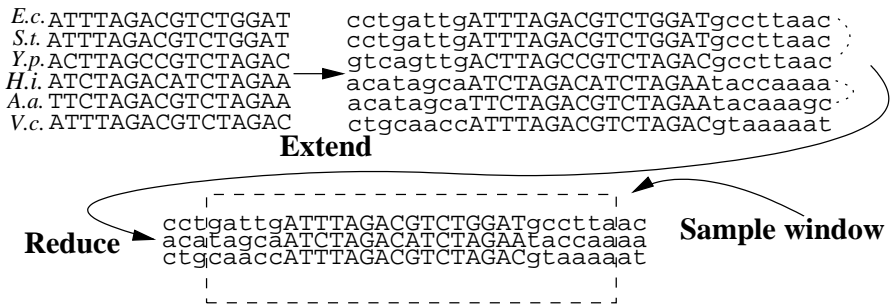,height=4cm}}
\caption{Operations on the data sets: Starting from an alignment of
variable length, we extend the alignment to length $32$ by
padding bases from the genome, and then replace sequences of
closely-related species by their consensus. This yields so-called
mini-WMs which are the objects that our algorithm clusters. When moved
between clusters, a window of length $27$ is sampled from the
alignment.}
\label{opseq_fig}
\end{figure}

Our primary data sets \cite{McCueEtAl2001,RajewskyEtAl2001} consist of
alignments of relatively short sequences, i.e. typically $15-25$
bases, that where extracted from upstream regions of orthologous genes
in different prokaryotic genomes. Data set \cite{McCueEtAl2001} uses
the genomes of {\it E. coli}, {\it A. actinomycetemcomitans}, {\it
H. influenzae}, {\it P. aeruginosa}, {\it S. putrefaciens}, {\it
S. typhi}, {\it T. ferrooxidans}, {\it V. cholerae}, and {\it
Y. pestis}. Data set \cite{RajewskyEtAl2001} uses {\it E. coli}, {\it
K. pneumoniae}, {\it S. typhi}, {\it V. cholerae} and {\it
Y. pestis}. An example alignment is shown in the upper left of
Fig. \ref{opseq_fig}. The available evidence suggests that these
alignments either include or substantially overlap a set of binding
sites for a TF (or another kind of regulatory site). Our algorithm
will have to decide which stretch of bases in each alignment
corresponds to the regulatory site. Known binding sites \cite{church}
are between $11$ and $50$ bases long, with a mean of $24.5$ and a
standard deviation of just under $10$. We will assume that all binding
sites are exactly $27$ bases long, compromising between diluting the
signal in the small binding sites, and missing some of the signal in
long binding sites. We symmetrically expand the alignments in our data
set to length $32$, padding bases from the genomes (see
Fig. \ref{opseq_fig}).  We would like to treat these sequences as {\em
independent} samples of a single WM, but for closely related species,
this assumption is probably untenable. For alignments from data set
\cite{McCueEtAl2001} we therefore replace sites from the triplet {\it
E. c.}, {\it Y. p.}, and {\it S. t.}, and from the duplet {\it H. i.},
and {\it A. a.} by their respective consensi. For the data set
\cite{RajewskyEtAl2001} we only replace the triplet {\it E. c.}, {\it
K. p.}  and {\it S. t.} by their consensus.  The 'mini-WMs' so
obtained are the objects that our algorithm clusters. Finally, every
time the Monte-Carlo algorithm reassigns a mini-WM to a cluster, it
samples over the $6$ different ways of picking a length $27$ window
out of the length $32$ alignment, and over both strands (see
apppendix \ref{monte_carlo_sampling}).

Before clustering these primary data sets we tested the algorithm on a
set of experimentally determined TF binding sites in {\it E. coli}
that was collected in Ref. \cite{church}. We again extended (or
cropped) these sequences symmetrically to length 32. After excluding
$\sigma$ factor sites and sites that overlap one another by 27 or more
bases there are 397 binding sites representing 53 TFs remaining in
this test set. See appendix \ref{operations_on_the_data_sets} for
details on the pre-processing of this and our other data sets.

For the data set \cite{McCueEtAl2001} we removed all alignments that
overlap known binding sites or repetitive elements and then took the
top 2000 non-overlapping alignments ordered by MAP score. For data set
\cite{RajewskyEtAl2001} we also took the top 2000 non-overlapping
sites based on significance, but we left sites overlapping known
binding sites in this set. Finally, in order to separate new regulons
from new sites for TFs with sites in the collection \cite{church}, we
aligned all known {\it E. coli} sites for each TF into its own mini-WM
and added these 56 mini-WMs\footnote{3 out of the 53 TFs (argR, metJ,
and phoB) have two different types of sites which we align separately
into mini-WMs.} to the sets \cite{McCueEtAl2001} and
\cite{RajewskyEtAl2001}. Both these sets thus contain 2056 mini-WMs.

We created an additional test set consisting of the $397$ known
binding sites from \cite{church} and the E. coli sequences of the top
2000 unannotated mini-WMs from \cite{McCueEtAl2001}. As described
below, this test verified our prediction that by embedding the $397$
known sites in a larger set of sites, many clusters will fail to be
correctly inferred.

\section{Results}

We used the test-set of 397 known binding sites in several
ways. First, we sampled $P(C|D)$ and measured, for each factor, how
well its sites cluster. That is, we measured the co-clustering
distribution $p(k)$ for each TF. Using the significance threshold
described above, we found significant clusters for 24 of the 53
TFs. Twenty two TFs have 3 or fewer sites in the test set and, with
the exception of trpR, their sites did not cluster significantly. As a
better test of our algorithm, we compared the clusters inferred from
annealing this data set with the site annotation. We performed two
annealing runs to identify a ML partition, and then performed sampling
runs to test the significance of these maximum likelihood clusters. We
found that, in general, there is good agreement between the annotation
and the clusters inferred by annealing. For 17 of the 24 TFs that form
significant clusters there was an analogous significant cluster
obtained by the annealing. The full results are in appendix
\ref{res_clus_known_sites}. We have also found
that the likelihood $P(C|D)$ for the partition obtained in all
annealing runs is significantly {\em higher} than that obtained when
the sites are partitioned according to their annotation. Thus we feel
that the clustering for this data set can not be improved within our
scoring scheme. In short, our algorithm recovers almost half of all
regulons for which binding sites are known, and the large majority of
regulons for which there are more than three sites known.

We sampled $P(C|D)$ for the 2397 site test-set and found that, as
predicted, many clusters are lost (only 9 of 24 significant clusters
remain). Several of those that remain where reinforced by the presence
of additional unannotated sites in the supplemental set of 2000. (More
samples improves clusterability as we have seen in the clusterability
section.) For this larger data set, the total number of clusters
fluctuates around 350 during the run, but only $\sim5$\% of these are
significant. This suggests that most E. coli binding sites are in the
unclusterable regime, and that comparative genomic information is
essential to effectively cluster. We also performed simulations with
'surrogate' data sets that further support this claim. For each
cluster of known binding sites, we calculated the information score
$I$ of its WM and created 4 {\em random} WMs with equal $I$. By
drawing samples from each of these, we `scaled up' the set of known
binding sites and clusters by a factor of 5, to correspond to the
the estimated number of TF in E. coli. In sampling
$P(C|D)$ for this set, we found that less than 10\% of the clusters
are correctly inferred.

For the larger data sets from \cite{McCueEtAl2001,RajewskyEtAl2001},
that are our main interest, repeated annealing and sampling runs
indicated that both the annealed state and the significance statistics
are not fully converged within our running times ($10^{10}$ steps,
taking a week on a workstation per run). We therefore extracted
significant clusters via pair statistics as described above, which did
converge and allowed us to assign error-bars to all pair
statistics. For the data set \cite{McCueEtAl2001}, there were $365 \pm
5$ clusters on average and the connectivity graph gave 274 components
containing 1139 out of 2056 mini-WMs. Thus, about half of the data
set clusters stably, while the other half moves in and out of the
$\sim$100 unstable clusters. There were 115 significant clusters
comprising, 645 mini-WMs. Of the $115$ significant clusters, 21
contained, as one of its member mini-WMs, the alignment of a set of
known binding sites for the same TF from \cite{church}. These
clusters thus contain new sites for known regulons. The other 94
clusters correspond to new putative regulons some examples of which
are described below.

It is interesting to calculate the cluster information scores, $I$, to
compute the fractions, $e^{-I}$, of sequence space occupied by our
clusters. Summing these volumes, we find that approximately 1\% of the
space is filled by the top 45 clusters, the top 80 clusters fill 10\%
of the space, and all our 115 significant clusters fill 39\% of the
space. This again supports the idea that the set of all WMs is close
to the classification boundary: their binding sites fill almost
the entire sequence space.

For the data set \cite{RajewskyEtAl2001} there are $275 \pm 4$
clusters on average during the sampling. The connectivity graph has
176 clusters containing 726 mini-WMs. There were 65 significant
clusters (containing 398 mini-WMs), of which 25 correspond to known
regulons.  With respect to the sequence space volume filled by the WMs
of these clusters: 1\% of the space is filled by the first 30
clusters, 50 clusters fill 10\% of the space, and the full set of 65
WMs fills about 50\% of the sequence space.

\section{Examples}

Table \ref{wads_clus_table} contains a synopsis of some of predicted
new regulons we have examined in detail from the data set
\cite{McCueEtAl2001}. Primary cluster membership is noted along with
additional sites that can be found by scanning the cluster WM over the
full data set and all regulatory regions of E. coli. The complete
lists are on our web site \cite{website}.

\begin{table}[htbp]
\caption{Sample clusters from the data set \protect\cite{McCueEtAl2001}. The
cluster rank is by WM information score. The defining operons come in
three categories; Those with member sites in the data set on which the
algorithm was run (bold). Those with sites in \protect\cite{McCueEtAl2001}
that match the WM (normal font), and those that were found by scanning
the regulatory regions of E. coli (italics). Multiple genes within an
operon are separated by a / or by multiple capitals at the end of the
gene name. Operons separated by a comma indicates that the site fell
between divergently described genes.}
\begin{center}
{\scriptsize
\begin{tabular}{|c|c|c|} \cline{1-3}
cluster name & rank & defining operons \\
\hline
\hline
thiamin biosynthesis & 0 & {\bf thiCEFGH} {\bf tpbA/yabKJ} {\bf thiMD} {\it thiL} \\
\hline
gntR/idnR regulon & 1 & {\bf idnK,idnDOTR} {\bf gntKU} {\bf gntT} {\bf b2740} {\it edd/eda}\\
\hline
elongation factor & 2 & {\bf tufB}\\
\hline
ribonucleotide reductase & 3 & {\bf nrdAB} {\bf nrdDG} {\it nrdHIEF}\\
\hline
? & 4 & {\bf coaA} {\bf tgt/yajCD/secDF} {\bf yegQ} {\it b3975} {\it
tpr} {\it yeeO}\\
\hline
stem-loop/attenuator & 5 & {\bf yhbc/nusA/infB} {\bf mutM} {\bf
arsRBC} {\bf yhdNM}\\
repair ? & & {\bf nadA/pnuC} {\bf lig} ptsHI/crr rbfA/truB/rpsO\\
\hline
ntrC regulon & 11 & {\bf glnK/amtB cmk/rpsA glnALG glnHPQ}\\
& &  narGHJI hisJQMP\\
\hline
ribosomal protein attenuation & 15 & {\bf thdF} {\bf fabF} {\bf recQ} {\bf tsf}
{\bf pnp} {\bf pyrE} himD\\
\hline
anaerobic oxidation & 16 & {\bf cydAB} {\bf appCB} {\bf yhhK,livKHMGF}\\
& &  {\bf torCAD,torR} ansB/yggM {\it ybbQ} {\it yiiE} \\
\hline
fatty acid biosynthesis & 17 & {\bf fabA} {\bf b2899(yqfA)} {\it fabB} {\it
fabHDG}\\
\hline
cell envelope  & 25 & {\bf pcnB/folK pssA dksA/yadB yaeS}\\
replication ? & &  {\bf mreCD/yhdE/cafA} sanA {\it cmk/rpsA}\\
\hline
alkaline phosphatase & 26 & {\bf yaiB/phoA/psiF, ddlA} {\bf dnaB/alr}\\
peptidoglycan & & {\bf creABCD} iap avtA\\
\hline
transport & 37 & {\bf  abc,yaeD cadBA araFGH,yecI celABCDF}\\ 
& & {\bf citAB,citCDEF} {\it agaBCD tauABCD}\\ 
\hline
fruR regulon & 71 & {\bf fruR fruBKA epd yggR}\\
\hline
Fe-S radicals   & 85 & {\bf metK,yqgD  ftn pykA} yheA/bfr\\
\hline
\end{tabular}}
\end{center}
\label{wads_clus_table}
\end{table}

Our thiamin cluster is an example of a predicted new regulon that has
recently been experimentally confirmed. A comprehensive review of
thiamin biosynthesis in prokaryotes \cite{begley} places the genes
from the three operons of our thiamin cluster\footnote{thiBPQ is also
called tbpA/yabJK} into a single pathway, together with the four
single genes: thiL, thiK, dxs (yajP), and thiI (yajK). A recent paper
\cite{MirandaEtAl2001} shows that the three thiamin operons share a
common RNA stem-loop motif which is responsible for
posttranscriptional regulation. It is precisely a portion of this
motif that we cluster. A fragment of this structure also occurs just
upstream of translation start in thiL. For the remaining genes thiK,
dxs, and thiI, there are no putative sites in the data set
\cite{McCueEtAl2001}.

Besides the main gluconate metabolism pathway, a second pathway that
utilizes input from the catabolism of L-idonic acid has recently been
reported \cite{BauschEtAl1998} and corresponds to our second cluster.
The first two operons (idnK, idnDOTR) code for the enzymes that import
L-idonate and convert it to 6-P-gluconate.  The operon gntKU contains
a a gluconokinase, which catalyzes the same reaction as the idnK
protein, and a low-affinity gluconate permease. b2740 is a gene of
unknown function which belongs to the family of gluconate
transporters. Finally, gntT is a high affinity gluconate permease.
Additional sites were found upstream of the edd/eda operon, which
encode the key enzymes of the Entner-Doudoroff pathway
\cite{peekhaus1}. Ref \cite{BauschEtAl1998} suggests that idnR both
upregulates the L-idonate catabolism genes and represses gntKU and
gntT when growing on L-idonate. This suggest our sites may bind
indR. However, there are two sites upstream of gntT, which are
annotated as gntR sites \cite{peekhaus2}, which are also part of our cluster.

The pathway for ribonucleotide reduction to deoxyribonucleotides is
pictured on p. 591 of \cite{neidhardt} and includes the first two
operons of our like named cluster. We did not find sites in the
regulatory regions of the other two genes in this pathway (ndk,
dcd). Scanning of the genome with the WM inferred from the nrdAB and
nrdDG sites reveals an additional 3 (weaker) sites upstream of the
nrdHIEF operon. The nrdEF genes are annotated as a cryptic
ribonucleotide reductase. The regulation of our two primary operons
(nrdAB, nrdDG) is known to be complex and includes an fnr site
upstream of nrdD (which we correctly clustered with other fnr sites),
and additional fis, dnaA, and unattributed sites upstream of nrdA
\cite{Jacobson&Fuchs1998}. The nrdA site in our cluster is a new site,
located down stream of transcription start. Since nrdA is down
regulated during anaerobiosis, and nrdD is essential for anaerobic
growth, we would guess that our sites are involved in the switch.

The estimated WM of cluster 5 has a prominent inverted repeat sequence
as its consensus:\\ AAAAacCC***TT***GGGGgTTTTTT,\\ and has over 20
matches in the genome. These sites may correspond to an RNA secondary
structure, possibly involved in attenuation. There is no clear
predominant functional theme to the genes in our cluster 5. Noteworthy
are sites upstream of the arsenic resistance operon (arsRBC), the crr
regulator of a multidrug efflux pump, and the ydnM (zntR) regulator
for Pb(II), Cd(II), and Zn(II) efflux. Also, two genes involved in DNA
repair occur (MutM, lig).

The sites in cluster 15 occur upstream of genes whose proteins are
involved in RNA modification (thdF, pnp), recombination (recQ, himD)
and translation (tsf). More strikingly, 6 out of 7 of these sites
occur {\em downstream} of genes coding for ribosomal protein subunits
and one RNase. For 5 of these, there is evidence \cite{EcoCyc} that
our site falls within a transcription unit, i.e. that the genes
upstream and downstream of our site are co-transcribed. It seems
likely that these sites are involved in either attenuation or else
translational regulation.

E. coli has a rich repertory of respiratory chains that are built from
a variety of electron donors and acceptors, see p. 218
\cite{neidhardt}. One of our clusters (16) involves two homologous
cytochrome operons cydAB and appCB (cyxAB), which transfer electrons
to oxygen and are mainly active during anaerobic conditions. The
torACD operon (divergently transcribed with its regulator torR)
transfers electrons to trimethylamine N-oxide. There is a third
cytochrome complex, cyoABCD, with different specificity, that is not
linked to this cluster. Other operons in this cluster, such as
livKHMGF, which is involved in amino acid import, and ansB, which
catalyzes asparagine to aspartate conversion, seem unrelated but are
divergently transcribed with genes of unknown function. However,
Ref. \cite{jennings} and p. 366 \cite{neidhardt} suggest that ansB,
can also provide fumarate as a terminal electron acceptor. AnsB is
strongly upregulated during anaerobic conditions and has known crp and
fnr sites. The ansB site in our cluster is different from these
sites.

Cluster number 17 corresponds to the fatty acid biosynthesis regulon
with TF yijC (fabR) that was identified in
\cite{McCueEtAl2001}. Our cluster contains the sites they found
upstream of fabA and b2899. Additionally, we found WM matches upstream
of the related genes fabB and fabHDG. Other operons with lower quality
sites in the cluster include the mglBAC operon (methyl-galactoside
transport), clpX (component of clpP serine protease), and the putative
peptidase b2324.

We are unable to guess the functional role of the binding sites
clustered in cluster number 25. Some of the genes have functionalities
related to the cell envelope and membrane (pssA, yaeS, mreCD, sanA)
and some seem involved in replication (dskA, cafE). However, these
functions seem rather diverse.

For cluster 26, we find sites upstream of genes involved in
peptidoglycan biosynthesis (alr, ddlA, avtA, mrcB), and genes that are
known to be regulated in response to phosphate starvation (creABC,
iap, phoA/psiF). In particular, alkaline phosphatase (phoA) is
upregulated more than 1000-fold and accounts for as much as 6\% of the
protein content of the cell during phosphate starvation, see p. 1361
\cite{neidhardt}. Since alkaline phosphatase is active in the
periplasm, it seems conceivable that peptidoglycan synthesis is
down-regulated when phoA is expressed at such high levels.

Additional clusters with obvious common functionality include: cluster
85 for Fe-S radical synthesis \cite{cheek}, and the large cluster 37
which contains several PTS and other transport systems. Cluster 71
contains sites that overlap binding sites for the fructose repressor
fruR. These were clustered separately from the known fruR sites
because of a systematic shift, larger than the range our algorithm
scans, between how they were given in \cite{McCueEtAl2001} and the
annotated fruR sites. Similarly, cluster 11 contains sites that
overlap binding sites for the nitrogen fixation regulator ntrC (glnG).

Apart from these putative new regulons, our web site \cite{website}
has an additional 270 unique sites that cluster with WMs of known
TFs. Summing their membership probabilities, this corresponds to an
expected 135 new binding sites. The website also provides information
for each E. coli gene separately: inferred regulatory sites upstream
of the gene and the cluster memberships of these sites.

The clusters inferred from the data \cite{RajewskyEtAl2001} are also
on our web site \cite{website}. We have not yet evaluated their
functional significance, but some of them correspond
to clusters that we also found in the data of \cite{McCueEtAl2001},
e.g., the thiamin cluster reappears.

\section{Discussion}

We introduced a new inference procedure for probabilistically
partitioning a set of DNA sequences into clusters. Currently, the
algorithm assumes all WMs to be of a fixed length, but prior
information about site lengths, their dimeric nature, and the length
of spacers between dimeric sites could be easily included. One could
also extend the hypothesis space on which the algorithm operates: one
may assume that only some fraction, rather than all, of the sequences
are WM samples, while the rest should described by a `background'
model. This would, for instance, be appropriate for analyzing entire
upstream regions. In all these generalizations, the algorithm would
still assign probabilities to sets of sequences belonging to a single
TF. This essentially Bayesian approach should be contrasted with
approaches, e.g. \cite{bussemaker,consensus}, in which 'promising'
motifs are selected based on how {\em unlikely} it is for them to
occur under some null hypothesis of randomness.

By applying our algorithm to data sets
\cite{McCueEtAl2001,RajewskyEtAl2001} of putative regulatory sites
extracted from enteric bacteria, we predicted $\sim100$ new regulons
in E. coli, containing $\sim500$ new binding sites, and predicted
$\sim150$ new binding sites for known TFs. The functionality of many
of the predicted new regulons is supported by the fact that their
sites are found upstream of genes that are clearly functionally
related. Even if there is no common theme in the annotation of the
genes controlled by the sites, our significance measures suggest that
a large fraction of the clusters is functional: the data-sets contain
only conserved sites upstream of orthologous genes in different
organisms, {\em and} a highly significant association of groups of
such sites was found. We note that our set is a considerable
augmentation of the $\sim400$ non-$\sigma$ sites that are known
experimentally. Analysis of some our clusters shows that included in
our predicted regulons in addition to TF binding sites, are RNA stems
controlling translation, and even termination motifs.

The clusters and sites resulting from our genome wide analysis of
regulatory motifs allows for a more quantitative evaluation of the
global structure of regulatory networks in bacteria. The regulatory
network is often imagined as a rather loosely coupled collection of
'modules', where each regulon controls a set of genes with closely
linked functionality (although, many exceptions of course exist, such
as the structural TFs fis, ihf, etc.). Our predicted regulons are
often much less orderly. In several cases, some but not all genes of a
well studied pathway entered the regulon. In other cases, a regulon
contains sets of sites for genes of two or three clearly distinct
functionalities for which no regulatory connection is known. Our
overall impression is of a more haphazard regulatory network than
traditionally imagined.

Finally, we have emphasized the distinction between classifying and
clustering a set of binding sites. We have argued that the TFs of a
cell are essentially solving a classification task, and that inferring
regulons from the set of binding sites of a single genome may well be
impossible {\em in principle}. There are also evolutionary arguments
that support this claim. Like any piece of DNA, binding sites are
subject to random mutations. The more specific binding sites are, the
more likely they are to be disrupted by mutations. Evolution will thus
naturally drive TFs and their binding sites to become as unspecific as
possible \cite{NimwEtAl99,Borispaper} within the constraints set by
their function. That is, evolution will drive the set of binding sites
toward the 'classification threshold' where they become
unclusterable. The situation is reminiscent of the situation in
communication theory, where optimally coded messages look entirely
random to receivers that are not in possession of the
code. Information from comparative genomics is thus essential for the
inference of regulons from genomic data, and as the number of
sequenced genomes grows, so will our algorithm's ability to discover
new regulons.

\begin{acknowledgments}
The support of the NSF under grant number DMR-0129848 is acknowledged.
\end{acknowledgments}

\appendix

\section{The Weight Matrix Representation of Binding Sites}

In this section we briefly explain which assumptions are implied in
the weight matrix (WM) representation of transcription factor binding
sites. Transcription factors (TFs) will bind specifically to certain
DNA segments in the genome. We wish to mathematically represent the
sequence features, shared by these segments, that are responsible for
the specific recognition by the TF. It is generally
assumed \cite{von_hippel} that the binding energy $E(s)$ of the
TF to a DNA sequence segment $s$ can be written as a
sum of binding energies $E_i(s_i)$ for the different bases in the
segment:
\begin{equation}
E(s) = \sum_{i=1}^l E_i(s_i) 
\end{equation}

The probability $P_{\rm bound}(s)$ that the TF binds
to a sequence $s$ in the genome is then proportional to 
\begin{equation}
P_{\rm bound}(s) \propto e^{\beta E(s)},
\end{equation}
where $\beta = 1/(k T)$. We now further assume that binding sites
for a particular TF are distinguished from all other DNA segments in
that they have higher binding energies. That is, binding sites are
characterized by having some expected binding energy $\langle E
\rangle$ which is substantially higher than the expected binding
energy of random sequences. Under these assumptions, the probability
distribution $P(s)$ that segment $s$ is a binding site for the TF is
given by the maximum entropy distribution under the condition $\langle
E(s) \rangle = \langle E \rangle$,
\begin{equation}
P(s) = \frac{e^{\lambda E(s)}}{\sum_{s'} e^{\lambda E(s')}} =
\prod_{i=1}^l \frac{e^{\lambda E_i(s_i)}}{\sum_{\alpha} e^{\lambda
E_i(\alpha)}},
\end{equation}
where the sum over $s'$ is over all length-$l$ sequences, and the sum
over $\alpha$ is over the four bases. The Lagrangian multiplier
$\lambda$ is chosen such that $\sum_s E(s) P(s) = \langle E \rangle$.
If we define the WM
\begin{equation}
w^i_{\alpha} =  \frac{e^{\lambda E_i(\alpha)}}{\sum_{\alpha'} e^{\lambda
E_i(\alpha')}} ,
\end{equation}
then we can represent TFs by WMs, and write
for the probability $P(s|w)$ that $s$ is a binding site for the
TF represented by $w$ as 
\begin{equation}
P(s|w) = \prod_{i=1}^l w^i_{s_i}.
\end{equation}

In summary, under the assumption that the binding energy of a
TF to a DNA segment $s$ can be written as a sum of
the binding energies of the individual bases in $s$ to the
TF, and assuming that binding sites for a
TF can be characterized by having a certain expected
binding energy $\langle E \rangle$, we can represent TFs by WMs $w$. For each length-$l$ DNA segment $s$,
the probability of this segments being a binding site for
TF $w$ is simply given by a product of the WM probabilities of the individual bases, as described above.

\subsection{Information Scores}

If we have an alignment of $n$ length-$l$ sequences, with $n^i_\alpha$
occurrences of base $\alpha$ at position $i$, we can ask for the
probability $P(n^i_\alpha | b_\alpha)$ of  observing these base counts
under the assumption that all these sequences are random, with 
probabilities $b_\alpha$ for base $\alpha$ occurring at each
position. This probability is given by
\begin{equation}
\label{exact_expression_sample}
P(n^i_\alpha | b_\alpha) = \prod_{i=1}^l n! \prod_\alpha
\frac{(b_\alpha)^{n^i_\alpha}}{n^i_\alpha!}.
\end{equation}
If we use the Stirling approximation for the factorials (valid for $n$
moderately large):
\begin{equation}
\log(n!) \approx n \log(n) - n,
\end{equation}
and write for the WM entries
\begin{equation}
n^i_\alpha = n w^i_\alpha,
\end{equation}
we obtain
\begin{equation}
P(n^i_\alpha | b_\alpha) = \exp \left[-n \sum_{i,\alpha} w^i_\alpha
\log(w^i_\alpha/b_\alpha) \right] \equiv e^{-n I},
\end{equation}
where we have defined the WM information score $I$ as
\begin{equation}
\label{information_score_expression}
I = \sum_{i,\alpha} w^i_\alpha \log(w^i_\alpha/b_\alpha).
\end{equation}
Thus, the higher the information score of the observed alignment, the
less likely it is to occur by chance from sampling random bases. This
is the main reason that most authors use the information score to
assess the quality of alignments of putative binding sites. 

Obviously, for small $n^i_\alpha$ and especially if some of the
$n^i_\alpha$ become zero, the Stirling approximation breaks down, and
it would be best to use the exact (\ref{exact_expression_sample}). The
information score can thus best be {\em defined} as
\begin{equation}
I = -\frac{\log[P(n^i_\alpha | b_\alpha)]}{n}
\end{equation}
from the exact expression for all values of $n$. 

\section{The Partition Likelihood Function}

We recall that the probability $P(S|w)$ that all sequences of a set
$S$ were drawn from a WM $w$ is given by 
\begin{equation}
\label{p_set_from_w}
P(S|w) = \prod_{s \in S} \prod_{i=1}^l w^i_{s_i},
\end{equation}
where $s_i$ is the base occurring at position $i$ in sequence $s$.  To
calculate the probability $P(S)$ that all sequences in a set $S$
were drawn from the {\em same} WM, independent of what this
WM may be, we integrated over the hypersimplex
\begin{equation}
\sum_{\alpha} w^i_{\alpha} = 1, \; \forall i \in \lbrace 1, 2,\ldots
l \rbrace.
\end{equation}
Generally, one would have to choose some integration measure $\pi(w)$
on this space (which is equivalent to choosing a prior). We then have
\begin{equation}
P(S) = \int \, \pi(w) P(S|w) dw.
\end{equation}
We chose a uniform prior that normalizes the integrals
\begin{equation}
\pi(w) = (3!)^{l}.
\end{equation}
In cases where the sampling distribution is multinomial (as it is in
our case), it is customary to use so-called Dirichlet priors
\begin{equation}
\pi(w) = \prod_{i,\alpha} (w^i_{\alpha})^{\lambda_{\alpha}-1}.
\end{equation}
Demanding a prior that is invariant under scale transformations of the
$w^i_{\alpha}$, one would arrive at $\lambda_{\alpha} = 0$. There are
good arguments for suggesting that this corresponds to a complete {\em
ignorance} prior for $w$ \cite{Jaynescollection}. It is also easy
to derive that setting $\lambda_{\alpha}$ to some value larger than
zero is equivalent to adding $\lambda_{\alpha}$ counts for each letter
$\alpha$ to the data. For this reason, the $\lambda_\alpha$ are also
referred to as pseudo-counts. We interpret our use of $\lambda_\alpha
= 1$ as representing the knowledge that it is definitely {\em
possible} for each base $\alpha$ to occur at each position. That is,
we known $w^i_{\alpha} > 0$ for all $\alpha$. In any case, these are
subtleties that hardly affect the inference. For instance, some
authors (e.g., ref. \cite{Gibbssampler}) prefer to choose
$\lambda_\alpha$ proportional both to the frequency of $\alpha$ in the
``full'' data set (being for instance the genome or all of its
intergenic regions) and proportional to the square root of the number
of sequences in the alignment. We, however, fail to see how a
representation of our {\em prior} knowledge should depend on the data
set, nor why WMs should {\em a priori} be likely to be skewed toward
bases that have a higher frequency of occurrence in the full genome or
its upstream regions (which consist mostly of DNA that is {\em not}
part of a binding site). Still, we have experimented with such priors
and noticed that the clustering behavior is hardly affected at all. We
thus chose to use the uniform prior since it is simpler, and more
easily justifiable theoretically.

Finally, some remarks on the importance of the integration over $w$ in
calculating the probability of a cluster. First of all, it is of
course simply a matter of probability theory that if we want to
calculate the probability that all sequences in the cluster came from
the same WM, in absence of any knowledge about this WM, we have to
integrate over $w$. However, one may for instance be tempted to assign
a ``score'' to each cluster by finding the WM $w^*$ that maximizes the
likelihood $P(S|w^*)$ that all sequences in the cluster came from this
WM. One would find
\begin{equation}
P(S|w^*) = \prod_{i=1}^l \prod_{\alpha}
\left(\frac{n^i_{\alpha}}{n}\right)^{n^i_{\alpha}},
\end{equation}
where the maximum likelihood (ML) WM has entries
$(w^*)^i_{\alpha} = n^i_{\alpha}/n$.

The problem with an approach like this is that, trivially, the
partition in which each sequence forms its own cluster will be the
partition that maximizes the score. That is, for single-sequence
clusters one can always find a WM that gives rise to that sequence
with probability $1$. Obviously, we would not want this totally
``unclustered'' state to have the highest score. The source of the
problem is that to obtain these ML scores, one has to choose all WMs
such that the likelihood is maximized. The more clusters there are in
the partition, the more WMs have to be set to their ML values. One
thus allows oneself to optimize over more degrees of freedom for a
partition with many clusters than for a partition with a small number
of clusters. One then might be tempted to introduce some kind of
``penalty'' per degree of freedom that has to be optimized to obtain
the ML score. However, these problems are solved automatically when
one applies probability theory correctly, that is, when one performs
the integration over $w$.

\subsection{Estimating WM Entries}

Given an alignment of sequences that we believe to be sampled from the
same WM, we would like to estimate the WM entries $w^i_\alpha$ from which
these sequences were sampled. In general we have for the expected WM
entry $\langle w^i_\alpha \rangle$
\begin{equation}
\langle w^i_\alpha \rangle = \frac{\int w^i_\alpha P(S|w) \pi(w)
dw}{\int P(S|w) \pi(w) dw},
\end{equation}
where the integral is again over the hypersimplex, and $\pi(w)$ is
again the prior on this space. With the general Dirichlet prior one finds
\begin{equation}
\langle w^i_\alpha \rangle = \frac{n^i_{\alpha} + \lambda_\alpha}{\sum_{\beta}
n^i_\beta + \lambda_\beta}.
\end{equation}
With our prior $\lambda_\alpha =1$, we thus have $\langle w^i_\alpha
\rangle = (n^i_\alpha+1)/(n+4)$. 

\subsection{Relation to the Gibbs Sampler}

Consider a situation in which there only two clusters: one very large
``reservoir'' cluster and one small cluster (which contains the
estimated alignment of binding sites for a single TF). Under our
model, this partition has a certain probability $P$. We are now
interested in calculating the {\em change} $P \rightarrow P'$ in that
probability when a single sequence is moved out of the reservoir
cluster into the smaller cluster. With $n^i_{\alpha}$, the number of
bases $\alpha$ in column $i$ before the move and $N_\alpha$ the number
of bases $\alpha$ in the ``background reservoir'' (which is assumed to
have equal base frequencies $b_{\alpha}$ in each column), this is
given by
\begin{equation}
P' = P \prod_{i=1}^l \frac{n^i_{\alpha}+1}{n+4} \frac{N+3}{N_{\alpha}}
\equiv P Q.
\end{equation}
In the limit of $N$ goes to infinity the factor $Q$ reduces to
\begin{equation}
Q = \prod_{i=1}^l \frac{n^i_{\alpha}+1}{(n+4) b_{\alpha}} =
\prod_{i=1}^l \frac{\langle w^i_{\alpha} \rangle}{b_{\alpha}},
\end{equation}
with $b_{\alpha}$ the background base frequencies  and $\langle
w^i_{\alpha} \rangle$ the expected WM entries based on the
$n$ members of the smaller cluster prior to the addition of the new
sequence. This factor $Q$ is precisely (assuming uniform
pseudo-counts) the scoring that is used by the Gibbs sampler
algorithm. Thus, under the assumption that the data set consists of a
large set of ``background'' sequences plus a set of binding sites for a
single TF, our scoring reduces to the scoring of
Gibbs sampler. Note, however, that assuming a single cluster plus
background is of course not correct for data sets of the type that we
cluster in the paper.

\subsection{Prior on the Space of Partitions}

In the paper, we are using a uniform prior over all possible
partitions. The argument for using such a prior is that it is the
maximum entropy prior with respect to the space of partitions. In
absence of any prior knowledge about which partitions are more or less
likely, a uniform prior is the least ``assuming'' of any prior. We
note, however, that there are some subtleties with this reasoning.
Instead of saying that we are {\em a priori} completely ignorant
regarding which partition is more or less likely, it may seem that we
may just as well have said that we are completely ignorant regarding
the number of clusters that underlie the data. These two statements
are mutually inconsistent, however. The number of ways of partitioning
a set of $N$ objects into $n$ subsets is given by a so-called Stirling
number of the second kind $S^N_n$. The total number of ways of
partitioning a set of $N$ objects into subsets is
\begin{equation}
b_N \equiv \sum_{n=1}^N S^N_n.
\end{equation}
A uniform prior over the partitions thus corresponds to a probability
distribution
\begin{equation}
P_n = \frac{S^N_n}{b_N}
\end{equation}
over the number of clusters $n$. This distribution has, for large $N$,
a relatively sharp peak at some value of $n$. Therefore, complete
ignorance regarding the partition ``induces'' knowledge regarding the
number of clusters.

We could of course have chosen a uniform prior over the number of
clusters, thereby increasing the relative weight of partitions with
cluster numbers that can be realized in fewer ways. We are, however,
not trying to infer the cluster {\em number}, we are inferring which
particular partitions are most likely. We therefore feel that a
uniform prior of partitions is the most relevant for our purposes.

\section{Monte Carlo Sampling}
\label{monte_carlo_sampling}

In this section we explain in more detail how our Monte Carlo sampling
is implemented. First of all, we want to implement a move set that
would sample all partitions of our set of mini-WMs equally often if
the probability distribution $P(D|C)$ were {\em constant}. That is, we
want to implement a move set that respects the uniform prior. This can
be done as follows.

Let our data set contain $N$ mini-WMs and imagine that we have $N$
boxes. A clustering state can be specified by assigning the $N$
mini-WMs to the $N$ boxes (leaving some boxes empty of course). There
are $N^N$ such states. This state space (call it $X$) can be easily
uniformly sampled by, at each time step, picking one of the $N$
objects at rando, and moving it to one of the $N-1$ other boxes,
which is also to be chosen at random. However, a uniform sampling of
$X$ does not correspond to a uniform sampling of the space of
partitions. For each possible partition $C$ of the data set, there is
a multitude of states in the state space $X$ that correspond to this
partition $C$. To be precise, for a partition $C$ containing $n$
clusters there are $N!/(N-n)!$ states in the state space $X$ that
correspond to the same partition $C$ (all ways of permuting the $N$
boxes, divided by all ways of permuting the empty boxes). Therefore,
partitions with $n+1$ clusters have $N-n$ times as many states in $X$
as partitions with $n$ clusters. We thus bias the move-set on $X$ such
that the rate of moving from a state with $n+1$ clusters to a state
with $n$ clusters is $N-n$ times as high as the backward rate. This
will ensure a uniform sampling of the space of partitions.

A second issue that the Monte Carlo walk has to deal with is the
aligning of the mini-WMs with respect to each other. As
mentioned in the text, all our mini-WMs are $32$ bases
long, while we assume that the binding sites are $27$ bases
long. There are six ways of placing a length $27$ window over the
length $32$ mini-WM. Every time a mini-WM is
moved during the random walk, we sample over these six possible
shifts {\em and} over both strands of the DNA (leading to $12$
possible ways of picking the ``site'' from the length $32$ window). This
sampling is done by so-called importance sampling.

Assume that we want to consider moving mini-WM $m$ from a
cluster containing the set of sequences $S$ into a cluster containing
the set of sequences $S'$. Let $P(S')$ be the probability of the set
$S'$ without $m$ added, and let $P(S^-)$ be the probability of the set
of sequences $S$ when $m$ is {\em removed}. Further, let $P(S',i,s)$
be the probability of the cluster $S'$ when mini-WM $m$ is
added, with a shift $i$ and using strand $s$ (where $i$ runs from $0$
to $5$, and $s$ is either positive or reverse complement). Finally,
let $P(S^-,i,s)$ be the probability of cluster $S$ with $m$ added at
shift $i$ and strand $s$. We then define
\begin{equation}
Z = P(S')\sum_{i=0}^5 \sum_{s \in \{+,-\}} P(S^-,i,s),
\end{equation}
and
\begin{equation}
Z' = P(S^-) \sum_{i=0}^5 \sum_{s \in \{+,-\}} P(S',i,s).
\end{equation}
The move of $m$ from $S$ to $S'$ is accepted when $Z' \geq Z$, and
accepted with probability $Z'/Z$ when $Z' < Z$. After $m$ is assigned
to either $S$ or $S'$, one of the $12$ strand/shift combinations is
sampled with probability $P(S^-,i,s)/Z$ (or $P(S',i,s)/Z'$ depending on
which cluster $m$ was assigned to). This extension of the move set will
sample over all possible shift and strand combinations that the
clusters can take.

With this move set, it still seems that stable clusters may get
``trapped'' into an unfavorable positioning of their window. That is,
assume that some subset of mini-WMs has high similarity and can form a
stable cluster. Let us also assume that these sites are not shifted
with respect to each other, so that with high likelihood an alignment
will be sampled in which all mini-WMs in the cluster occur with the
same shift. Still, one may choose different values for this shift,
i.e. the alignment may contain bases $0$-$26$, or bases $1$-$27$,
etc. Once one of these shifts is chosen our move set makes it very
unlikely for the cluster to revert to another shift unless it
evaporates. To defeat clusters getting trapped in a particular shift
this way we implemented one more move: the ``coherent shift''. In this
move, we sample, according to their probability, from all ways of
shifting all mini-WMs in the cluster by the same amount. This sampling
is performed (with probability $p$) on the cluster that contains the
mini-WM that was randomly selected to be reassigned. In our
implementation we chose $p = 0.05$. The behavior of the algorithm is
insensitive to this value. One only has to make sure that the coherent
shift is tried at least once during the lifetime of a typical cluster.

Finally, we altered the probability $P(S)$ for clusters consisting of
a single mini-WM. For those single mini-WM
clusters, we interpret the mini-WM as stemming from a
random background distribution as opposed to it consisting of samples
from a binding sites. That is, instead of a score $4^{l n}$ (with $l$
the length and $n$ the number of sequences in the mini-WM)
we use a probability
\begin{equation}
P = \prod_{s \in S} \prod_{i=1}^l b_{s_i},
\end{equation}
where $S$ is the set of sequences that make up the mini-WM, and
$b_{\alpha}$ is the back ground frequency of letter $\alpha$ in the
upstream regions. It must be admitted that this is somewhat of a
hybrid procedure. If we had {\em really} wanted to consider a
hypothesis space in which some of the mini-WMs can be ``background''
sequences as opposed to WM samples, then we should have chosen a prior
on the space of partitions that properly reflects this.

A typical Monte Carlo run has on the order of $10^{10}$ (considered)
moves. It takes less than $10^9$ moves to reach ``equilibrium''. That
is, for the first $10^8 - 10^9$ time steps the sampled values of
$P(D|C)$ are decreasing, whereas they fluctuate around some fixed
value after $10^9$ time steps. For the data set from
ref. \cite{McCueEtAl2001} this equilibrium value is given by
\begin{equation}
\log[P_{\rm ss}] \approx 0.93 \log[P_{\rm uc}],
\end{equation}
where $P_{\rm ss}$ is the steady-state value of the probability
$P(D|C)$ and $P_{\rm uc}$ is the probability of the entirely
unclustered state where each mini-WM forms a cluster by
itself.

\subsection{Annealing}

The annealing is performed by raising all probabilities to the power
$\beta$ and slowly raising $\beta$ over time. That is, a move form
partition $C$ to $C'$ is accepted with probability
$[P(D|C')/P(D|C)]^{\beta}$. 

We experimented with different annealing schedules, and a simple
linear increase of $\beta$ with time gives the best results. We let
the algorithm run at $\beta = 1$ for $10^8$ steps and then start
increasing $\beta$ linearly with time, such that $\beta \approx 3$ in
the end phase of the annealing. The last $5 * 10^7$ time steps are
then done with a very high value of $\beta$, essentially only
accepting moves that increase the probability $P(D|C)$, thereby
leading to a locally optimal partition at the end of the run.

\section{Extracting Significant Clusters}
\label{extract_sig_clusters}

The simplest way in which we could use our assignment of probabilities
to partitions to extract clusters is by treating the probabilities
simply as a ``scoring'' function and attempting to find a partition of
optimal score. One could for instance use hierarchical clustering:
\begin{enumerate}
\item{Start with each mini-WM forming its own cluster.}
\item{Calculate, for all pairs of clusters, the change in the
probability of the partition when these clusters are
fused (maximizing for each pair over all possible shift and strand combinations).}
\item{When there is no fusion that increases the partition's
probability: stop. Else, fuse those clusters that increase the
partition's probability most and go to step 2.}
\end{enumerate}
Experiments with the test set of $397$ known TF binding sites
\cite{church})show that (choosing cut offs as favorably as
possible) no more than $10$ regulons are correctly inferred using this
procedure. Moreover, such greedy-search algorithms only search a
minute portion of the state space. But most importantly, such schemes
ignore one of the key advantages of our method, which is that it
automatically assigns probabilities to clusters.

\subsection{Tracking Stable Clusters}

One way of identifying significant clusters from the Monte Carlo
sampling is to look for clusters that are long-lived. A cluster is
``born'' when a mini-WM is assigned to form a pair with a mini-WM that
is forming a cluster by itself. We then follow this cluster until it
``evaporates'', which occurs when only a single mini-WM is left. We
record which mini-WMs were part of this cluster during its
lifetime. That is, we compile a probabilistic membership distribution
$p$, where $p_i$ is the percentage of the lifetime of the cluster
that mini-WM $i$ was assigned to this cluster. We only keep track of
such lifetime membership distributions for clusters whose lifetimes
are over some threshold (short transient formations of pairs, for
instance, are not recorded).

At the end of the Monte Carlo run, we will have a large collection of
such clusters $c$, with their lifetimes $T_c$, and their membership
distributions $p^c$. There generally will be sets of clusters that
have very similar membership distributions. We want to consider such
sets of similar clusters as different samples of the {\em same}
cluster. That is, we imagine that there is some set of ``real'' and
``unique'' clusters, and that the clusters that we recorded during the
Monte Carlo random walk are {\em samples} from this set of underlying
clusters. We introduce a probability $P(p^c,p^{c'})$ that clusters $c$
and $c'$ are samples from the same underlying cluster. We imagine that
each underlying cluster has some intrinsic membership distribution $p$
and that our example clusters $p^c$ and $p^{c'}$ are samples of size
$T_c$ and $T_{c'}$ of this intrinsic membership distribution. The
probability $P(p^c,p^{c'})$ then can be derived analogously to the way
in which we derived the probability $P(S)$ for all sequences in a set
$S$ to come from the same WM.

For each component $i$, we have $n^c_i = T_c p^c_i$ for the number of
times that mini-WM $i$ occurred in cluster $c$. Now given the
intrinsic membership probability $p_i$ we would have
\begin{equation}
P(n^c_i,n^{c'}_i | p_i) = (p_i)^{n^c_i+n^{c'}_i}
(1-p_i)^{T_c+T_{c'}-n^c_i-n^{c'}_i}.
\end{equation}
Because we do not known the value of $p_i$, we again integrate over it
\begin{equation}
P(n^c_i,n^{c'}_i) = \int_{0}^1 \, P(n^c_i,n^{c'}_i | p_i) d p_i =
\frac{ (n^c_i+n^{c'}_i)!
(T_c+T_{c'}-n^c_i-n^{c'}_i)!}{(T_c+T_{c'}+1)!}
\end{equation}
Similarly, for the probability that these two samples did {\em not}
come from the same distribution, we have
\begin{equation}
P(n^c_i)P(n^{c'}_i) = \frac{n^c_i! n^{c'}_i! (T_c-n^c_i)!
(T_{c'}-n^{c'}_i)!}{(T_c+1)! (T_{c'}+1)!}.
\end{equation}
Therefore, the increase in the probability $P \rightarrow P'$ when
clusters $c$ and $c'$ are fused, is given by
\begin{equation}
P' = P \prod_{i=1}^N \frac{P(n^c_i,n^{c'}_i)}{P(n^c_i)P(n^{c'}_i)}.
\end{equation}

We have implemented a hierarchical clustering that starts out assuming
that all recorded clusters are unique and then iteratively fuses
those clusters for which the increase in probability is highest. We
then cut this procedure off at a more or less arbitrarily chosen
point. 

We found this procedure (although workable and giving similar results
for the significant clusters) ultimately unsatisfactory because there
is no natural way to cut off the hierarchical clustering, and because
this is again a greedy algorithm that only searches a small part of
the space. In a sense, we end up with the same clustering problem all
over again. We sampled the distribution $P(C|D)$ and obtained a set of
clusters. However, since these clusters are clearly not unique, we
essentially have to cluster this set of clusters to obtain unique
ones! (We could of course sample again all ways of fusing the
clusters... and then end up with probabilistic clusters of
clusters... and so on {\em ad infinitum}).

The key conceptual difficulty here is that there is no clear
definition of a {\em probabilistic cluster}. A partition is a rigid
assignment of objects into groups, and since our probability
assignment gives an essentially fluid assignment of objects into groups,
it it not clear how to extract ``clusters'' from this fluid assignment.

It is therefore that we chose to simply identify the ML rigid
assignment of the mini-WMs into clusters by annealing. At the end of
the annealing, we have a rigid assignment of mini-WMs into groups. Our
Monte Carlo sampling is then used to estimate the significance of
these clusters. This is currently the most satisfying approach that we
have for identifying significant clusters, and it is the one we used
on the test set of 397 known binding sites from ref. \cite{church}.

\subsection{Estimating Significance of ML clusters}
\label{Est_Sig_of_ML_Clus}

The annealing essentially gives us a set of ``candidate'' clusters
without determining the significance of these clusters. We only know
that a search for the ML partition has partitioned the data into these
clusters. We measure the significance of these candidate clusters by a
Monte Carlo sampling run.

Let $M$ be the set of mini-WMs making up a particular candidate
cluster. At each particular time step during the Monte Carlo sampling
this set of mini-WMs will be partitioned over a certain number of
clusters. Some members may occur together in a cluster, while other
members may be partitioned with some other mini-WMs. In general, the
set of members $M$ of the candidate cluster will be spread over a set
of clusters $c$, that each contain $m_c$ members of the set $M$ (with
$\sum_c m_c = |M|$, of course). We now find the maximum $m_{*} = {\rm
max}_c(m_c)$ of these values $m_c$ and say that at this particular
time step $m_{*}$ members of the candidate cluster $M$ are
``coclustered'' (that is, we find the cluster that contains most of
the members of $M$ and count how many members of $M$ are in this
cluster). By recording $m_{*}$ at each time step, we calculate a
distribution $p(m)$ for the probability of $m$ members of $M$
coclustering. We can then of course calculate mean, variance and so
forth of this distribution. In particular we will calculate the
shortest interval $[m_{-},m_+]$ that contains $95$\% of the
probability: $\sum_{m = m_-}^{m_+} p(m) \geq 0.95$. If $m_{-} > 1$ we
consider the cluster to be ``significant''.

We can also measure, for each member of $M$, what fraction of the time
$p$ it is a member of the set of $m_{*}$ members. We then obtain
a probabilistic membership list for this cluster. That is, for each
member $m$ we obtain the probability $p_m$ that $m$ is ``in'' the
cluster.

\subsection{Extracting Significant Clusters from Pairwise Statistics}

For larger data sets it may be computationally unfeasible to converge
all of the statistics just mentioned. We may then need to resort to
measuring a simpler set of statistics to infer significant clusters. 

\begin{figure}[htbp]
\centerline{\epsfig{file=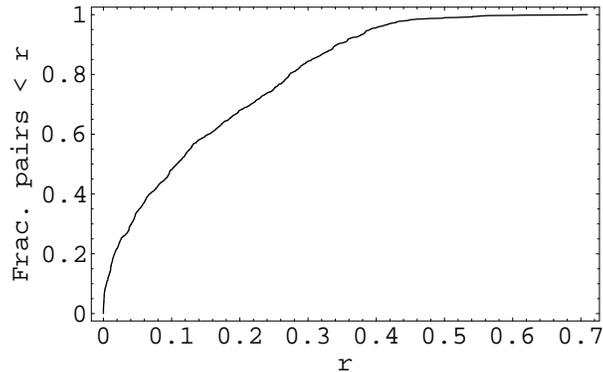,height=5cm}}
\caption{Cumulative distribution of the ratio $r$ of standard deviation to
mean for all pair statistics with a mean larger than $1/2$.}
\label{pair_conv_fig}
\end{figure}

Our approach is to measure, for each pair of mini-WMs $i$ and $j$, the
fraction of the time $p_{ij}$ that these mini-WMs are coclustered,
i.e. occur in the same cluster. We do several ($20$ or so) Monte Carlo
runs and measure all $p_{ij}$ in each of them. By comparing the
measured $p_{ij}$ from the different runs we can obtain both the mean
and the standard deviation between different runs for each $p_{ij}$
(giving us a direct handle on the convergence). For a data set of $N$
mini-WMs, we thus measure mean $\mu$ and standard
deviation $\sigma$ for each of the $N(N-1)/2$ pair statistics
$p_{ij}$. In order to illustrate the convergence attained in our runs
Fig. \ref{pair_conv_fig} shows the cumulative distribution of the
ratio $r= \sigma/\mu$ of the standard deviation and mean for all pairs
that have a mean of over half, $\mu > 1/2$. (We focus on these because
these are the pair statistics that will be used below.)
The figure shows, for instance, that $50$\% of the pairs have a standard
deviation smaller than $10$\% of the mean. 

\begin{figure}[htbp]
\centerline{\epsfig{file=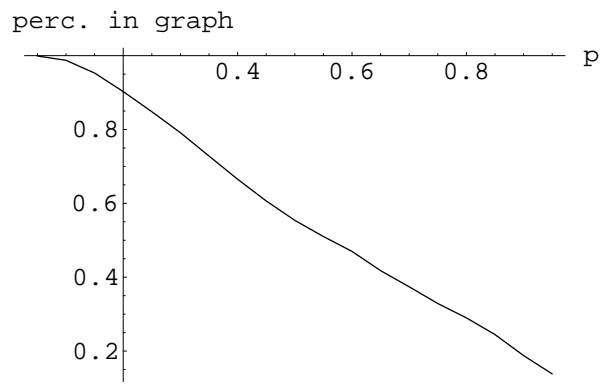,height=5cm}}
\caption{Fraction of the data set that has at least one other mini-WM
with which it clusters more than a fraction $p$ of the time. The
data set consists of  the top $2,000$ sites from ref. \protect\cite{McCueEtAl2001}.}
\label{frac_in_graph_fig}
\end{figure}

\begin{figure}[htbp]
\centerline{\epsfig{file=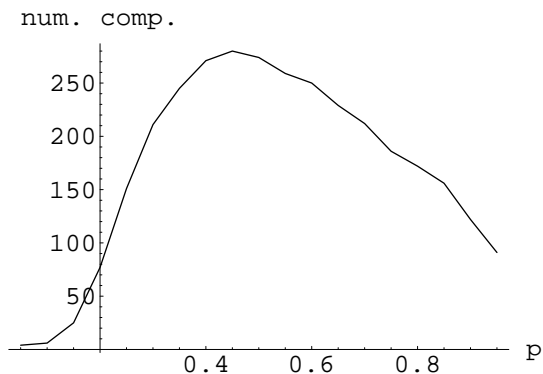,height=5cm}}
\caption{Number of connected components (with more than $1$ member) as
a function of the connectivity threshold $p$. The data set consists
of the top $2000$ sites from ref. \protect\cite{McCueEtAl2001}.}
\label{num_compdatanu_in_graph_fig}
\end{figure}

We now use the pair statistics to define ``candidate'' clusters. We
construct a graph where each node represents a mini-WM and connect
those nodes $i$ and $j$ for which $p_{ij} > p_{\rm crit}$, where
$p_{\rm crit}$ is some connectivity threshold. This graph will contain
a certain number of connected components. Some of these components
will consist of single nodes. These unconnected mini-WMs are
``orphans'' in the sense that there are no other mini-WMs with which
they cluster more than $p_{\rm crit}$ of the time. All other mini-WMs
have at least one partner with which they cluster more than a fraction
$p_{\rm crit}$ of the time. Fig. \ref{frac_in_graph_fig} shows the
fraction of the data set that has at least one partner over $p_{\rm
crit}$ as a function of $p_{\rm crit}$.
Fig. \ref{num_compdatanu_in_graph_fig} shows the number of connected
components, i.e. the number of candidate clusters, as a function of
$p_{\rm crit}$.
We see that when the threshold $p_{\rm crit}$ is set very high, for
instance at $95$\%, that there are slightly less than $100$ connected
components, containing somewhere between $10$ and $20$\% of the data
set. As the threshold $p_{\rm crit}$ is lowered, many new components
appear, and the fraction of the data set contained in these components
grows along with it. At some point (around $p_{\rm crit} = 0.45$), the
number of components reaches a maximum. For lower thresholds the
number of components drops quickly. What happens here is that the
different clusters start fusing rapidly for lower thresholds.

We chose to set the threshold at $p_{\rm crit} = 1/2$. One reason for
setting the threshold at this value is that it is conceptually most
appealing: a mini-WM belongs to a cluster if there is a mini-WM within
that cluster with which it coclusters {\em most} of the time. By
setting the threshold at $1/2$ we are guaranteed that each mini-WM
cannot cluster more with any other mini-WM than with those in the
cluster that it is assigned to. A second reason is pragmatic: Fig.
\ref{num_compdatanu_in_graph_fig} shows that around this value of the
threshold the number of different components is maximal, i.e. we will
obtain a maximal number of candidate clusters.

Now that we have our candidate clusters, we want to calculate their
significance by calculating the distribution $p(m)$ of the number of
its coclustering members (see section \ref{Est_Sig_of_ML_Clus}). We
also want to assign, to each of its members $m$, a membership
probability $p_m$. Both of these we want to infer from the pairwise
statistics (since it is computationally prohibitive to measure
all these statistics directly by sampling). 

The coclustering probabilities $p(m)$ should in principle be inferred
by first estimating the probability of coclustering for all possible
subsets of the candidate cluster. We would for instance need to
estimate, from the pair statistics, the probability $p_{ijk}$ that
members $i$, $j$, and $k$ all three co-occur in a single cluster. That
is, we need to estimate higher order coclustering statistics from the
pair statistics, i.e. given $p_{ij}$, $p_{ik}$, and $p_{jk}$ we want
to estimate $p_{ijk}$. This is, in principle, a well defined
probability theory problem. We have equalities such as
\begin{equation}
p_{ij} = p_{ijk} + p_{ij,k},
\end{equation}
where $p_{ij,k}$ is the probability that $i$ and $j$ but not $k$
occur together in a cluster. There are two analogous equalities for
$p_{ik}$ and $p_{jk}$. Finally, the probability  $p_{i,j,k}$ for all three to occur
in separate clusters has to be nonnegative, $p_{i,j,k} \geq 0$,
and the sum over all possible ways of partitioning $i$, $j$, and
$k$ should of course be $1$:
\begin{equation}
p_{ijk}+p_{ij,k}+p_{i,jk}+p_{ik,j}+p_{i,j,k} = 1.
\end{equation}
Given no other information than the pair statistics and the above
constraints one then finds a uniform distribution over $p_{ijk}$
within the interval bounded by ${\rm
max}(\{(p_{ij}+p_{ik}+p_{jk}-1)/2,0\})$ from below and ${\rm min}(\{ p_{ij} ,p_{ik}
,p_{jk}\})$ from
above. The expected value $\langle p_{ijk} \rangle$ would then be
given by
\begin{equation}
\langle p_{ijk} \rangle = \frac{{\rm min}(\{ p_{ij} ,p_{ik}
,p_{jk}\}) +{\rm max}(\{(p_{ij}+p_{ik}+p_{jk}-1)/2,0\})}{2}.
\end{equation}
In principle, all higher order statistics could be estimated in this
way. However, this sort of inference quickly becomes very cumbersome
as more indices are involved.

Instead, we estimate membership probabilities by a simpler
(approximate) procedure. We assume that one of the members of the
connected component, can be considered the ``anchoring'' or central
member of the cluster and that the probability for a mini-WM to be a
member of the cluster is the probability that it coclusters with the
anchoring member. We first calculate the probabilities $A_i$ that
mini-WM $i$ is the anchoring member. Then, given the $A_i$, the
cluster membership probability $p_i$ of mini-WM $i$ is given by
\begin{equation}
\label{membership_prob_eq}
p_i = A_i + \sum_{j \neq i} p_{ij} A_j.
\end{equation}
That is, either $i$ is the anchoring member itself, or $j$ is the
anchoring member and $i$ coclusters with it.
We now define the $A_i$ to be the normalized solution to
\begin{equation}
A_i = \lambda \sum_j p_{ij} A_j.
\end{equation}
The idea is that the probability that $i$ is the central member of the
cluster, should be proportional to the probability that $i$ coclusters
with the central member. 

We thus solve the above eigenvector equation for each connected
component of the graph, obtain the anchor probabilities $A_i$ and
calculate the membership probabilities for all its nodes using
(\ref{membership_prob_eq}).

Finally, we now use the membership probabilities to calculate the
probabilities $p(k)$ that $k$ members from a connected component
cocluster at any point in time. To this end, we will assume that for
each member $m$, the probability $p_m$ of this member being part of
the cluster, is {\em independent} of the probability of any of the
other members being part of the cluster. That is, each member $m$ has
an independent probability $p_m$ to be ``in'' the cluster. If we then
define the generating function
\begin{equation}
G(z) = \prod_m (p_m z + 1-p_m) = \sum_k p(k) z^k,
\end{equation}
the probabilities $p(k)$ can be easily obtained by expanding $G(z)$,
and we can again calculate the $95$\% probability interval
$[k_-,k_+]$. Significance of the cluster again is defined by $k_- >
1$.

\subsection{Estimating a Cluster's WM}

We now want, for each cluster, to estimate a WM from the
membership probabilities $p_m$ of its members $m$. The approach is
that if $m$ is in the cluster with probability $p_m$, that, with the
same probability $p_m$, the cluster WM is given by the
alignment of sites that $m$ belongs to. Therefore, we want to
calculate the average alignment of sites that each member $m$ belongs
to. Thus, during the Monte Carlo sampling, we keep track, for {\em
each} mini-WM $i$, of the alignment of the mini-WMs of the cluster in
which $i$ finds itself. That is, at any point during the run, mini-WM
$i$ will find itself in some cluster $c$, containing some set of other
mini-WMs. The alignment of the cluster is simply described by the
numbers of $n^i_{\alpha}$ of bases $\alpha$ at column $i$ of the
cluster. We will now keep track, for each mini-WM, of the running
average of this alignment. We of course have to take into account that
the mini-WMs under consideration may occur in the cluster with a
different length $27$ window sampled at different time steps. Thus if
at a certain point in time, the mini-WM occurs in the cluster with the
window $s$ through $s+27$, then we will add the base counts
$n^i_{\alpha}$ of the current cluster to columns $i+s$ of the mini-WMs
WM running average. In this way, we get an average
alignment for each of the $32$ columns of the mini-WM.

At the end of the sampling, we thus have, for each mini-WM, the
averaged alignment of the sites in the clusters that it visited during
the run. We will now reconstruct the WM of each ``candidate'' cluster
(be it a ML cluster, or one inferred from pair statistics) by summing
the averaged alignments of all members $m$ of the cluster, each
weighted with their membership probability $p_m$. To meaningfully do
this sum, we still have to align the averaged alignments of the
different cluster members with respect to each other. To this end, we
start with the averaged alignment of the member with the largest
membership probability $p_m$. We then align the averaged alignment of
the member with the second highest membership probability to this
first member. After that, we align the member with the third highest
membership probability to this pair, and so on, until all members have
been added to the alignment. For each column we then obtain averaged
base counts $\langle n^i_{\alpha}\rangle$. We finally obtain the
WM estimates from these averaged base counts:
\begin{equation}
\langle w^i_{\alpha} \rangle = \frac{ \langle n^i_{\alpha} \rangle
+1}{\langle n \rangle +4}
\end{equation}

\subsection{Membership Based on WM Match}

For each cluster that occurred at the end of annealing (in the
annealing approach) or that forms a connected component of the
pairwise clustering graph (in the approach via pair statistics) we
reconstruct the WM as described in the previous section. We now {\em
classify} the full data set of mini-WMs in terms of these estimated
cluster WMs. Let $P(S|w_j)$ be the probability that the sequences in
mini-WM $S$ arise from sampling from the cluster WM $w_j$. The
probability $P(w_j|S_i)$ that mini-WM $i$ was sampled from WM $w_j$,
as opposed to any of the other cluster WMs, is given by:
\begin{equation}
P(w_j|S_i) = \frac{P(S_i|w_j) P(w_j)}{\sum_k P(S_i|w_k) P(w_k)},
\end{equation}
where the $P(w_k)$ are the prior probabilities of cluster WMs, and the
sum in the denominator is over all cluster WMs (from either the set of
ML WMs or the set of connected components of the connectivity graph.)
With respect to the prior, we chose to use the prior that maximizes
the likelihood of the full data set $D$ given the cluster WMs. That
is, choose $P(w_j)$ such that
\begin{equation}
P(D) = \prod_{S \in D} \left[ \sum_j P(S|w_j) P(w_j) \right]
\end{equation}
is maximized, where the product is over all mini-WMs in the data
set. Once we have determined the prior, we obtain the posterior
probabilities of Eq. {\bf 39} and in this way obtain
another membership list for each cluster, but now for the full data
set. For each significant cluster these membership probabilities are
also shown on the website (www.physics.rockefeller.edu/$\sim$erik/website.html).

\subsection{Finding Matches in Upstream Regions of a Genome}

We can search for additional members of our clusters by scanning the
upstream regions of all operons in {\em E. coli} for matches to the cluster
WMs. For each cluster WM, and each upstream region $U$, we will
calculate the probability that this upstream region contains {\em no}
sites for the cluster WM. To calculate this, consider first some
particular length 27 sequence $s$ in the upstream region $U$. The
probability that $s$ arose from WM $w$ is $P(s|w)$ (given
by the usual expression). The probability $P(s|r)$ that this sequence
arises ``by chance'' is $P(s|r) = \prod_i b_{s_i}$, where $b_{\alpha}$
is the background frequency of base $\alpha$. Therefore, the
probability that sequence $s$ is a binding site rather than a
``random'' sequence is given by the posterior
\begin{equation}
P(w|s) = \frac{P(s|w) \pi}{P(s|w) \pi + (1-\pi) P(s|r)},
\end{equation}
where $\pi$ is the a prior probability that a particular segment of
the upstream region $U$ is a binding site for a particular TF. We will
choose this prior so as to reflect the guess that, on average, an
upstream region contains about $2$-$3$ binding sites (a conservative
estimate). Since there are $\approx 300$ different TFs, the
probability that an upstream region has a binding site for a particular
TF is, on average, about $1/100$. Furthermore, we would like to
bias the prior such that binding sites are more likely to be found
close to the translation start. We will therefore let the prior depend
on the distance $d$ from the sequence $s$ to the translation
start. That is, we take $\pi(d) = C/d$, where the constant $C$ is
chosen such that $\sum_d \pi(d) = 1/100$ over the whole upstream
region. Finally, to obtain the probability that the upstream region
contains {\em no} binding sites for the TF, we take the product
over $P(w|s)$ for all ways of picking a length $27$ sequence from the
upstream region, which gives the probability $\bar{P}$ that the
upstream region contains no site for $w$:
\begin{equation}
\bar{P} = \prod_{s \in U} P(w|s).
\end{equation}
On the website we report, for each significant cluster, the 35 operons
with the highest scores $-\log[\bar{P}]$ for their upstream regions.

\section{Resampling Test for Identifying Significant Clusters}

In the paper, we assay the significance of our clusters within a
Bayesian framework: given our set of putative sites we find clusters
of sites that are more likely to cluster with each other than to
cluster in any other combination with other sites. That is, we
calculate how likely it is that certain sets of sites belong together
relative to these sites clustering with other sites. In contrast, it
is quite customary to assay the significance of a cluster by comparing
some statistic that measures its ``quality'' against the typical
quality of clusters obtained by clustering ``random'' sequences. That
is, one assumes some null hypothesis reflecting ``randomness'' and
calculates how {\em unlikely} it is for a cluster of a certain quality
to arise from such random data. For instance, to test the significance
of a cluster of $n$ sequences having a WM score $I$ that was obtained
from clustering a data set of $N$ sequences in total, one may ask: how
likely is it that a set of $n$ sequences can be found, in a set of $N$
random sequences, that has a WM score at least as high as $I$?

There are several problems with such an approach. First of all, one
must be very careful when interpreting such a test. Rejecting the null
hypothesis of randomness means no more than: the null hypothesis does
not explain the data well. What kind of other hypotheses {\em would}
explain the data better is not addressed by such a test. That is, for
our problem, one can not logically make the step: since the null
hypothesis is rejected, the cluster must correspond to a ``real''
regulon. To make that inference one has to introduce a model that
explicitly calculates the probability to obtain certain clusters of
sites given that they {\em are} binding sites from the same regulon
(as we have done).

A more technical problem is posed by the choice of null
hypothesis. Often null hypotheses are obtained by randomizing the data
in some way, and then re-running the algorithm with the randomized
data. In such ``resampling'' tests, it is important that the
randomization only removes correlations or biases from the data that
are relevant for the inference at hand. That is, we only want to
remove biases in the data that are due to the occurrence of real
binding sites in the data. There may be other, irrelevant biases in
the data (such as the relative frequencies of the different bases)
that should {\em not} be removed by randomizing the data. Otherwise,
the rejection of the null hypothesis may only imply that the null
hypothesis does not reflect certain inherent biases in the data that
have nothing to do with the occurrence of binding sites. In general,
this is a hard problem; there are probably all kinds of biases in real
DNA sequences that have nothing to do with the occurrence of binding
sites in these sequences.

In spite of these problems, we did perform some resampling
significance test for our clusters. Note also that in section
\ref{res_clus_known_sites} we evaluate the results on a test set of
known binding sites, which allows us to explicitly calculate the rate
of false positives and false negatives for that test set. For the
resampling test, we constructed a randomized data set by permuting the
columns of each of the mini-WMs in our data set independently. This
retains the distribution of bases in all the individual columns of the
mini-WMs but removes any correlations in the order of these columns
between different mini-WMs, which is a conservative randomization
procedure.

\begin{figure}[htbp]
\centerline{\epsfig{file=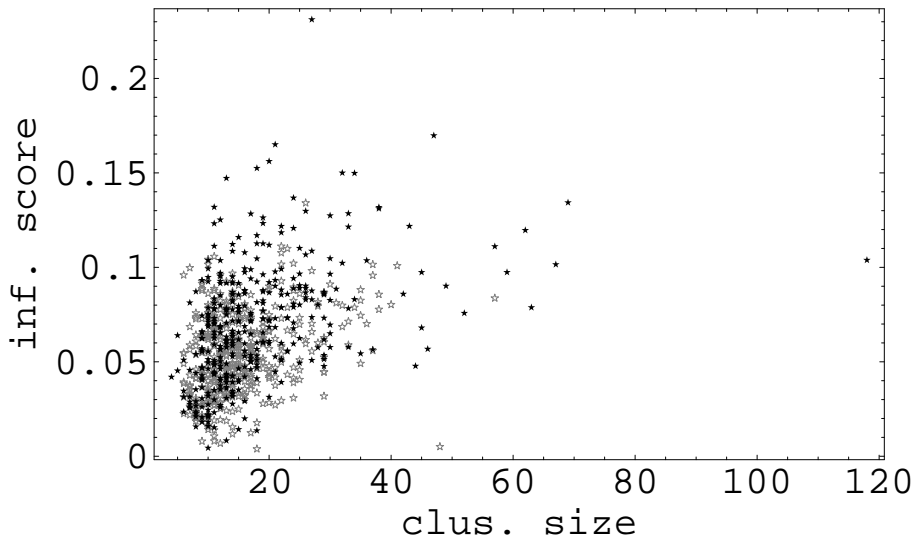,height=7.5cm}}
\caption{Resampling test for the clusters of data set from ref.
\protect\cite{McCueEtAl2001}. Each filled black star is a cluster obtained by
annealing the real data set, while each gray open star is a cluster
obtained by annealing the randomized data set. Cluster sizes are shown
horizontally, while their WM information scores $I$ are shown
vertically.}
\label{resam_test_fig}
\end{figure}

Fig. \ref{resam_test_fig} shows the results of the resampling test on
the results of data from ref. \cite{McCueEtAl2001}. The filled
black stars show the clusters that were obtained at the end of
annealing for the real data set. The number of sequences in each
cluster is shown on the horizontal axis, while each cluster's WM score
$I$ is shown on the vertical axis. Similarly, the gray open stars show
the clusters obtained by the annealing of the randomized data set. We
will call a cluster significant if there is no cluster of the same
size and with higher information score in the randomized data set. Of
course, it would be better to perform many resampling runs and
estimate more precisely the probability of annealing random data
producing a cluster with information score larger than $I$ at each
possible cluster size. This is computationally rather intensive. Since
we are assessing significance of clusters within a Bayesian framework,
we provide the data in Fig. \ref{resam_test_fig} as a guideline for
comparison with significance tests that are more commonly used. Under
our definition of significance for the resampling test we find $88$
significant clusters, which are mostly the same clusters as deemed
significant by our Bayesian procedure.

\section{Operations on the Data Sets}
\label{operations_on_the_data_sets}

In this section we briefly describe the data sets that were used in
our studies and describe in more detail the operations that we
performed on these data sets before clustering them using PROCSE.

The data set from ref. \cite{McCueEtAl2001} contains multiple
alignments of putative binding sites from a number of related
proteobacteria and was collected in the following way. For each
{\em E. coli} gene, orthologous genes were collected from the species
{\em Actinobacillus actinomycetemcomitans}, {\em Haemophilus}
influenzae, {\em Pseudomonas aeruginosa},
{\em Shewanella putrefaciens}, {\em Salmonella typhi}, {\em
Thiobacillus ferrooxidans}, {\em Vibrio cholerae}, and
{\em Yersinia pestis}. A modified version\footnote{One of the pertinent
modifications is the favoring of sites that show palindromicity.} of
the Gibbs sampler algorithm \cite{Gibbssampler} is then used to find
common motifs in the upstream regions of these orthologues. The
resulting alignments of putative binding sites contain between 1 and
16 sequence segments with lengths of $15$-$25$ bases. We will interpret
these alignments as collections of binding sites for the {\em same}
TF.

The data set from ref. \cite{RajewskyEtAl2001} contains similar
alignments of putative binding sites for a single TF but was obtained
in a somewhat different manner. Again, sets of orthologous genes were
collected for each {\em E. coli} gene from the species {\em Klebsiella
pneumoniae}, {\em S. typhi}, {\em V. cholerae}, and {\em
Y. pestis}. Instead of looking for similar sequence motifs in the
upstream regions, ref. \cite{RajewskyEtAl2001} aligned the upstream
regions and found sequence segments that are more conserved than
sequences downstream of orthologous genes. The resulting alignments
differ from those of ref.  \cite{McCueEtAl2001} in two
respects. First, there are more stringent constraints on the selection
of conserved sequence segments. Second, since the number of species
used is smaller, the number of sequences in the alignment per upstream
region is generally smaller.

\subsection{Operations on the Data Set from \cite{McCueEtAl2001}}

Ref. \cite{McCueEtAl2001} have annotated alignments in their data
set for which the {\em E. coli} site overlaps either a known site, or
a known repetitive element. We have removed all such alignments from
our data set, since for this data set, we are interested in clustering
new putative binding sites. We then ordered all the remaining
alignments by their MAP value. The MAP value is a quality measure
given to the alignments by ref. \cite{McCueEtAl2001}. We then went
through the ordered list of alignments and removed alignments whose
centers are less than $6$ bases away from centers of previous
alignments in the list. This is to avoid the possibility of the
algorithm aligning the same mini-WM {\em twice} in a cluster. Overlaps
are only checked for the E. coli members of the alignment. We then
took the top 2,000 of the remaining alignments as our data set.

After that we symmetrically extended all alignments from their
original length (between 15 and 25 bases) to length 32. To do so we
had to ``pad'' bases to the alignments from the genomes. Some of the
genomes used by ref. \cite{McCueEtAl2001} are not yet completed and
some cases occurred in which the bases could not be padded because the
sequence is not available. In those rare cases, we added random bases
to replace the missing bases. As pointed out in the paper, we would
like to consider the sequences in the resulting alignment as {\em
independent} samples from a WM describing the TF. However, some of the
species are probably too closely related evolutionary to warrant this
assumption of independence. That is, some of the bases in the
alignment are conserved not because of selectional constraints but
because they simply haven't mutated yet since the species
diverged. Ideally we would deal with these evolutionary relationships
directly, reconstructing the phylogenetic tree of the species
involved, and inferring which bases in the alignment have been
retained by selection. However, this is an altogether separate
inference problem and we opted for a simple circumvention of this
problem. We replaced the sequences from the most closely related
species by their consensus. That is, we replaced triplets of sequences
from {\em E. coli}, {\em S. typhi}, and {\em Y. pestis} with their
consensus, and replaced sequences from the pair {\em H. influenzae}
and {\em A. actinomycetemcomitans} with their consensus as well. We
deem all other species to be sufficiently separated evolutionary such
that we can assume their sequences to be independent samples of a
WM. There are two more technical details to mention regarding this
procedure. First, there may not be a majority base among the
sequences. In those cases we select one of the bases among the triplet
(or pair) of sequences at random. The second problem arises because
the alignments of ref. \cite{McCueEtAl2001} may contain more than
one sequence segment from the same upstream region, i.e. from the same
genome. Thus, if we have two segments from {\em S. typhi} and two from
{\em E. coli} occurring in the same alignment, we have to decide how to pair
these before taking their consensus. We do this by picking the pairing
that maximizes the number of conserved bases between the paired
segments. These operations now give us the mini-WMs on which the
PROCSE algorithm operates.

\subsection{Operation on the data set from \cite{RajewskyEtAl2001}}

The operations on the data set from ref. \cite{RajewskyEtAl2001}
are largely the same as those described in the previous section. We
used the set of predicted {\em E. coli} binding sites from
\cite{RajewskyEtAl2001} and the corresponding local alignments to
construct a mini-WM of length $32$ for each putative binding site
$x$. In case the length $l(x)$ was $<32$, we padded additional bases
to the left and to the right using the most significantly conserved
{\em E. coli} base $b$ as an anchor. Similarly, if $l(x)>32$, sites
were clipped off.  The sequences for {\em E. coli}, {\em
K. pneumoniae}, and {\em S. typhi} were replaced by their
consensus. Each mini-WM was scored by the significance score for $b$
\cite{RajewskyEtAl2001} and ordered accordingly. Alignments that
overlapped an alignment with higher score with $27$ or more bases were
removed. In contrast to the algorithm employed in
\cite{McCueEtAl2001}, the selection of sites in this data set was
not optimized to maximize the inclusion of known binding sites or to
favor reverse complement symmetry of the sites. For this reason, we
did not remove sites that overlap known binding sites (or known
repetitive elements) from this set. By comparing the results obtained
with this data set and those obtained with the data set from
\cite{McCueEtAl2001}, we may (to some extent) assess the effect on
the discovery of new vs. known regulons of the removal/inclusion of
known sites.

\subsection{Processing of the data set from \cite{church}}

The data set of known TF binding sites from Church's lab
\cite{church} contains a total of $997$ sites. In order to create a
test set for our algorithm we filtered this set in several ways. First
we removed all $\sigma$-factor binding sites. There are several
reasons for removing these sites. First of all, $\sigma$-$70$ sites
are ubiquitous in the genome, whereas we are interested in finding
regulons (sites and their TFs) that specifically regulate (relatively)
small sets of genes. Second, there are regulatory sites that overlap
$\sigma$-sites, and since in our algorithm any piece of genome can
only occur in one cluster at a time, these sites would alternatively
cluster with the regulon of interest and with the rest of the
$\sigma$-sites at other times. The presence of the $\sigma$-sites
would thus negatively effect the inference of other (smaller)
regulons. Finally, the primary data set \cite{McCueEtAl2001} looked
for reverse complement symmetry in the binding sites, thus disfavoring
the inclusion of $\sigma$-sites in their data set. Thus, removing the
$\sigma$-sites from the tes -set makes it also more comparable with
the data set from ref. \cite{McCueEtAl2001}.

After removing $\sigma$-sites and sites that overlap each other by
more than $27$ bases there are $397$ unique TF binding sites in this
data set.

\subsection{Preparation of TF mini-WMs from Data Set \cite{church}}

When we cluster the new putative binding sites we want to separate
clusters of sites that correspond to TFs for which some binding sites
are already known from clusters of sites that form entirely new
putative regulons. To this end we took all the known binding sites
from the collection from \cite{church} and divided them into groups
according to their annotation. That is, all sites that are annotated
as binding sites for the same TF form a group. We then aligned these
groups of sites into alignments of all known sites for each TF. The
multiple alignment is performed by running a hierarchical clustering
with our scoring function. That is, we first put all known sites for a
TF into separate clusters and then find the pair that, when fused,
leads to the largest increase in the score $P(D|C)$ of the
partition. We repeat this procedure (i.e. at the second step either a
third sequence is added to the pair or another pair is fused) until a
single cluster is left. At each step the alignment is maximized over
all possible ways of shifting the sites with respect to each other and
both strands. Thus, we sometimes have to pad bases from the genome to
``complete'' the alignment. At the end, the length $32$ window with the
highest information score is taken from the alignment. The mini-WM so
obtained represents all known sites for a particular TF. We
constructed $56$ different mini-WMs in this way from data set
\cite{church}. There are only $53$ TFs represented in this data set,
but three of them (metJ, argR, and phoB) have two different types of
sites, which we aligned separately.

We added these $56$ mini-WMs to both data sets. The idea is that if
additional binding sites for any one of these TFs occur in our data
set, then these sites will tend to cluster with one of the $56$ mini-WMs
that we have just described. When we collect the significant clusters
at the end of the run, we separate them into two sets: those that
contain at least one of the 56 mini-WMs constructed from \cite{church}
and those that do not. The former clusters contain new putative sites
for known regulons, while the latter contain new predicted regulons.

\section{Results of Clustering Known Binding Sites}
\label{res_clus_known_sites}

Table \ref{church_clus_table} shows results from clustering the set of
$397$ sites from ref. \cite{church}. We performed several different
tests of our algorithm with this data. 
\begin{table}[htbp]
{\scriptsize
\begin{center}
\begin{tabular}{|c|c|c||c|c|c||c|c|c|} \cline{1-9}
fac. & size & int. & size & num. site & int. & size
& num. site & int. \\
\cline{1-9}
arcA & 12 & 2-5 & 20/15 & 3/3 & 1-11/1-11 & 24 & 4 & 1-14 \\
\cline{1-9}
argR2 & 7 & 5-7 & 21/24 & 7/7 & 8-20/11-21 & 26 & 5 & 1-11 \\
\cline{1-9}
argR & 17 & 8-14 & 22/22 & 10/12 & 12-21/11-21 & - & - & - \\ \cline{1-9}
crp & 41 & 13-27 & 22/29 & 19/20 & 2-16/7-21 & 39 & 19 & 7-28 \\
\cline{1-9}
dnaA & 8 & 3-6 & 7/8 & 6/6 & 3-7/3-7 & - & - & - \\ \cline{1-9}
fadR & 7 & 4-7 & 12/9  & 7/7 & 3-10/3-8& 11 & 4 & 1-8 \\ \cline{1-9}
fis & 11 & 2-3 & 20/15 & 4/4 & 1-11/1-11 & - & - & - \\ \cline{1-9}
fnr & 12 & 7-9 & 11/11  & 9/9 & 7-10/7-10 & 25 & 6 & 17-23 \\
\cline{1-9}
fruR & 11 & 8-11 & 17/13 & 11/10 & 8-13/7-12 & 15 & 7 & 1-6 \\
\cline{1-9}
fur & 8 & 5-6 & 21/24 & 5/6 & 8-20/11-21 & 28 & 6 & 19-24 \\
\cline{1-9}
galR & 6 & 3-6 & 6/14 & 5/6 & 3-6/3-8 & 9 & 2 & 1-4 \\ \cline{1-9}
hipB & 4 & 3-4 & 8/14 & 3/4 & 2-5/4-13 & - & - & - \\ \cline{1-9}
ihf & 18 & 3-5 & 22/22 & 5/4 & 12-21/11-21 & 27 & 5 & 5-15 \\ \cline{1-9}
lexA & 19 & 18-19 & 24/24 & 19/19 & 19-23/19-23 & 28 & 18 & 19-24 \\
\cline{1-9}
lrp & 14 & 2-4 & -/- & -/- & -/- & - & - & - \\ \cline{1-9}
malT & 10 & 2-6 & 9/8 & 7/6 & 1-6/1-6 & - & - & - \\ \cline{1-9}
metJ(3) & 15 & 13-15 & 17/15 & 14/13 & 14-17/13-15 & 17 & 14 & 13-16 \\
\cline{1-9}
narL & 9 & 2-6 & 10/14 & 5/6 & 1-5/4-13 & - & - & - \\ \cline{1-9}
ntrC & 5 & 5-5 & 9/10 & 5/5 & 5-7/5-8 & 18 & 14 & 9-14 \\ \cline{1-9}
phoB & 9 & 2-4 & 9/13 & 3/4 & 1-7/1-8 & 11 & 4 & 1-8 \\ \cline{1-9}
purR & 16 & 13-15 & 15/17 & 14/16 & 13-15/13-16 & 23 & 13 & 16-20 \\
\cline{1-9}
trpR & 3 & 2-3 & 8/- & 3/- & 1-5/- & - & - & - \\ \cline{1-9}
tus & 5 & 4-5 & 11/8 & 5/5 & 5-8/5-7 & 12 & 5 & 3-6 \\ \cline{1-9}
tyrR & 16 & 5-11 & 15/21 & 9/8 & 2-10/1-13 & - & - & - \\ \cline{1-9}
\end{tabular}
\end{center}}
\caption{ Clustering of the $397$ sites from
ref. \protect\cite{church}. Column $1$ shows the TF name, column $2$ shows
the number of sites in the data set annotated as binding sites for
that TF (according to ref. \protect\cite{church}), column 3 shows the the
$90$\% probability interval for the number of coclustering sites for
this TF. Only those TFs are shown for which the lower bound of the
$90$\% probability interval is $2$ or higher. We performed two
annealing runs of this data set to find the ML partition. For each TF,
we found the cluster in this partition with the largest number of
sites for that TF. The clusters shown in columns $4$-$6$ resulted from
two simulated annealing runs, with the ``size'' and ``int'' defined as
before.  Column 5 shows how many of the sites in column 4 were
annotated for the TF. For example, columns $4$-$6$ for the TF dnaA
show that at the end of the two annealing runs there were clusters of
sizes $7$ and $8$, respectively that both contained $6$ of the $8$
known dnaA binding sites, and with $90$\% probability between $3$
and $7$ members of these clusters co-cluster. Finally, columns $7$-$9$ show
the same statistics when the annealing is done on a larger data set
that contains both the $397$ sites of ref. \protect\cite{church} as well as the
top $2,000$ E. coli sites from ref. \protect\cite{McCueEtAl2001}.}
\label{church_clus_table}
\end{table}
First of all, one may test how well the known binding sites cluster
according to their annotation. That is, do sites annotated to be
binding sites for the same TF $X$ significantly
cluster with each other?  To answer this question, we performed
Monte Carlo sampling of the probability distribution $P(C|D)$ for this
data set. At each time step of the Monte Carlo random walk, we measure
how many sites of each TF ``cocluster''. This is defined as follows: At
any particular time step of the Monte Carlo walk, the sites that are
annotated for TF $X$ will be distributed in some way among the
clusters that are present at that time during the random walk. For
instance, if TF $X$ has $10$ binding sites in the data, $6$ of
those may sit in one cluster, $2$ others may sit in another cluster
together with some sites of different annotation, and the last $2$ may
each sit in some cluster that is dominated by sites of other
annotation. For such a partition, we would say that $6$ of the $10$
sites for TF $X$ cocluster. By calculating how many sites
cocluster at each time step, we measure, for each TF, the
distribution $p(k)$ that $k$ of its sites will cocluster at any point
in time. These distributions may be summarized in various ways, but we
chose to find the shortest length interval $[k_{\rm min},k_{\rm max}]$
such that at least $90$\% of the time between $k_{\rm min}$ and
$k_{\rm max}$ sites cocluster, i.e. $\sum_{k=k_{\rm min}}^{k_{\max}}
p(k) > 0.9$. We will consider the sites of a TF $X$ to cluster
``significantly'' if $k_{\rm min} > 1$ for its distribution $p(k)$.  Of
the $53$ TFs represented in the data set, there are 24 that
cluster significantly. These are shown in the rows of Table
\ref{church_clus_table}. The first column shows the name of the
TF, the second column shows the total number of sites for that
TF contained in the $397$-site data set, and the third column
shows the interval $[k_{\rm min},k_{\rm max}]$. The fifth row, for
instance, shows that the TF dnaA has a total of $8$ sites in the
data set, and that at least $90$\% of the time between $3$ and $6$ of
these sites cocluster.

As mentioned before, $24$ of the $53$ TFs cluster significantly. One
might call this an overall ``false negative'' rate of $55$\%. All the
TFs that cluster significantly (with the exception of trpR) have more
than three binding sites in the data set. There are $22$ TFs that have
$3$ or less sites in the data. Therefore, it is clear that it is
mostly TFs with a small number of samples in the data that do not
cluster significantly. In other words, there is a false negative rate
of less than $23$\% for TFs with more than three sites in the data
set. Table \ref{church_non_sig} shows the TFs whose sites did not
cluster significantly. Note that some of these sites (such as cpxR,
flhCD, glpR, metR, modE, nagC, ompR) {\em do} cluster significantly
better than sets of random sequences would, but did not pass our
significance threshold\footnote{One would assume that for random data
all partitions are equally likely. One can then ask: what is the
expected co-clustering distribution $p(k)$ for some particular set of
$n$ sites, and what is the sampling distribution of, for instance, the
mean size $\langle k \rangle = \sum_k k p(k)$. Although cumbersome,
these quantities can be calculated analytically, and one would find
that for $n=6$, $\langle k \rangle = 1.39$ and that $\langle k \rangle
= 2$ constitutes a significant clustering of the $6$ sites. By such
measures, some of the clusters in table \ref{church_non_sig} would
come out significant.}. That is, we could have changed the significance
thresholds such that {\em for this particular data set} the number of
false negatives would be further decreased. Still, it is quite clear
that the sites for a TF like soxS do not cluster at all. The
variability among the $8$ soxS sites is so large that one might wonder
if a WM model is appropriate at all for TFs like this.

In this first test, we used the annotation of the sites to measure to
what extent sites with equal annotation cluster together. Summarizing,
we found significant clustering for slightly less than half of the
TFs, and for over $75$\% of the TFs that have more than 3 sites in the
data set. After this, we wanted to test to what extent our annealing
approach will be able to identify the ``correct'' clusters. That is,
we wanted to compare the ML clusters that result from annealing with
the annotation of the sites. We performed two annealing runs to
identify a ML partition. We tested the significance of the ML clusters
obtained by annealing by performing two sampling runs. Again we
measure the coclustering distribution $p(k)$ for each cluster and
consider a cluster significant if the $90$\% probability interval
$[k_{\rm min},k_{\rm max}]$ has $k_{\rm min} > 1$.
\begin{table}[htbp]
{\scriptsize
\begin{center}
\begin{tabular}{|c|c|c|} \cline{1-3}
fac. & size & int. \\
\cline{1-3}
ada & 2 &  1-1\\
\cline{1-3}
araC & 4 & 1-2\\ \cline{1-3}
carP & 2 & 1-1\\ \cline{1-3} 
cpxR & 6 & 1-3\\ \cline{1-3}
cspA & 4 & 1-2\\ \cline{1-3}
cynR & 2 & 1-1\\ \cline{1-3}
cysB & 3 & 1-1\\ \cline{1-3}
cytR & 5 & 1-2\\ \cline{1-3}
deoR & 1 & 1-1\\ \cline{1-3}
farR & 3 & 1-2\\ \cline{1-3}
fhlA & 3 & 1-1\\ \cline{1-3}
flhCD & 3 & 1-3\\ \cline{1-3}
gcvA & 4 & 1-2\\ \cline{1-3}
glpR & 7 & 1-3\\ \cline{1-3}
hns & 3 & 1-1\\ \cline{1-3}
hu & 1 & 1-1\\ \cline{1-3}
iclR & 1 & 1-1\\ \cline{1-3}
ilvY & 2 & 1-1\\ \cline{1-3}
lacI & 1 & 1-1\\ \cline{1-3}
marR & 2 & 1-2\\ \cline{1-3}
metR & 7 & 1-4\\ \cline{1-3}
modE & 3 & 1-3\\ \cline{1-3}
nagC & 6 & 1-3\\ \cline{1-3}
ompR & 8 & 1-3\\ \cline{1-3}
oxyR & 3 & 1-2\\ \cline{1-3}
pdhR & 2 & 1-1\\ \cline{1-3}
rhaS & 2 & 1-1\\ \cline{1-3}
soxS & 8 & 1-2\\ \cline{1-3}
torR &  1 & 1-3\\ \cline{1-3}
\end{tabular}
\end{center}}
\caption{TFs from data set \protect\cite{church} whose sites did not cluster
significantly. First column is TF name, second is the total number
of binding sites for that TF in the data, and the third column is
the $90$\% probability interval $[k_{\rm min},k_{\rm max}]$ of its
coclustering distribution $p(k)$.}
\label{church_non_sig}
\end{table}
In order to compare these clusters with the annotation, we had to
identify, for each TF, which of the ML clusters ``corresponds'' best
to the cluster of binding sites of the TF. We do this by simply
finding the ML cluster that contains the largest
number of sites for the TF. That is, the $10$ sites of TF $X$
may be spread over $4$ clusters in the ML partition, with $5$ sites in
one cluster, $3$ in another, and $2$ more in a third and fourth
cluster. We would then identify the ML cluster containing the $5$
sites as the cluster ``corresponding'' to TF $X$.

These clusters are shown in columns $4$ through $6$ of table
\ref{church_clus_table}. The column entries are best explained by
example. Focus, for instance, on the line for TF narL. In the
first annealing run, there was a ML cluster that contained $5$ of the
$9$ narL sites. This cluster had a total size of $10$ sequences (that
is, apart from the $5$ narL members, $5$ members had different
annotations). Under sampling, between $1$ and $5$ sites of this
cluster cocluster $90$\% of the time. (Note that this means: between
$1$ and $5$ of the full set of $10$ sites.) In the second annealing
run, there was a ML cluster of size $14$ that contains $6$ of the $9$
narL sites. Between $4$ and $13$ members of this cluster cocluster
$90$\% of the time.  To give an example of a TF for which
essentially the same cluster was found in both annealing runs, focus
on the lexA line. In both runs there was a cluster of size $24$ that
contained all $19$ of the known lexA sites (and $5$ sites with a
different annotation). During sampling, between $19$ and $23$ members
of this cluster cocluster (being one of the most stable
clusters in the set). 

In general, one can see that there is good agreement between the
annotation and the clusters inferred by annealing.  Looking in more
detail at some of the annealed clusters, we note that narL and narP
sites cluster together, as do fur and argR2 sites. The ihf sequences
sit mostly in the tail of the argR cluster and the fis sequences
occupy the tail of the arcA cluster. The many crp sites sit
distributed over two clusters (with only the largest represented in
the table).

This test also allows us to assess the amount of {\em false positives}
that our algorithm produces. At the end of the annealing there were
$30$ and $29$ clusters in the two respective runs. After sampling, it
turns out that $20$ of the $30$ we deemed significant using our
cut-offs, and $16$ of the $29$ in the other run. None of these
clusters are false positives. That is, our false positive rate is $0$
for this particular test set. 

Finally, one may think that the annealing had not really found the
ML clustering (or something close to that), but had
gotten stuck in some local optimum, and that if we had had more time,
and cooled slower, we may have found clusterings with better score
that correspond even better to the known annotation. This is
unlikely. One can easily calculate what the probability $P(C|D)$ is of
the partition in which all sites are clustered according to their
annotation. We find that this probability is substantially {\em lower}
than the probability of the partitions that annealing finds.  We
believe that this has two main reasons. First, we believe that some of
the sites are mis-annotated. Some sites of annotation $X$ consistently
cluster with sites of some other annotation $Y$. Second, for some
TFs (such as lrp, fis, ihf and soxS mentioned above) it seems that
the WM representation of the sites is unsatisfactory. Their WM scores
are extremely low: when we, for instance, run the clustering algorithm
on {\em only} the fis sites, the algorithm prefers partitions in which
these sites are not partitioned into a single cluster at all!
Possibly, these TFs recognize something else than a sequence motif
that can be represented by a WM.

\bibliography{/home/golem/erik/Ecoli/Notes/epev}

\begin{thebibliography}{29}
\expandafter\ifx\csname natexlab\endcsname\relax\def\natexlab#1{#1}\fi
\expandafter\ifx\csname bibnamefont\endcsname\relax
  \def\bibnamefont#1{#1}\fi
\expandafter\ifx\csname bibfnamefont\endcsname\relax
  \def\bibfnamefont#1{#1}\fi
\expandafter\ifx\csname citenamefont\endcsname\relax
  \def\citenamefont#1{#1}\fi
\expandafter\ifx\csname url\endcsname\relax
  \def\url#1{\texttt{#1}}\fi
\expandafter\ifx\csname urlprefix\endcsname\relax\def\urlprefix{URL }\fi
\providecommand{\bibinfo}[2]{#2}
\providecommand{\eprint}[2][]{\url{#2}}

\bibitem[{\citenamefont{McCue et~al.}(2001)\citenamefont{McCue, Thompson,
  Carmack, Ryan, Liu, Derbyshire, and Lawrence}}]{McCueEtAl2001}
\bibinfo{author}{\bibfnamefont{L.~A.} \bibnamefont{McCue}},
  \bibinfo{author}{\bibfnamefont{W.}~\bibnamefont{Thompson}},
  \bibinfo{author}{\bibfnamefont{C.~S.} \bibnamefont{Carmack}},
  \bibinfo{author}{\bibfnamefont{M.~P.} \bibnamefont{Ryan}},
  \bibinfo{author}{\bibfnamefont{J.~S.} \bibnamefont{Liu}},
  \bibinfo{author}{\bibfnamefont{V.}~\bibnamefont{Derbyshire}},
  \bibnamefont{and} \bibinfo{author}{\bibfnamefont{C.~E.}
  \bibnamefont{Lawrence}}, \bibinfo{journal}{Nucleic Acids Research}
  \textbf{\bibinfo{volume}{29}}, \bibinfo{pages}{774} (\bibinfo{year}{2001}).

\bibitem[{\citenamefont{Rajewsky et~al.}(2002)\citenamefont{Rajewsky, Socci,
  Zapotocky, and Siggia}}]{RajewskyEtAl2001}
\bibinfo{author}{\bibfnamefont{N.}~\bibnamefont{Rajewsky}},
  \bibinfo{author}{\bibfnamefont{N.~D.} \bibnamefont{Socci}},
  \bibinfo{author}{\bibfnamefont{M.}~\bibnamefont{Zapotocky}},
  \bibnamefont{and} \bibinfo{author}{\bibfnamefont{E.~D.}
  \bibnamefont{Siggia}}, \bibinfo{journal}{Genome Research}
  \textbf{\bibinfo{volume}{12}}, \bibinfo{pages}{298} (\bibinfo{year}{2002}).

\bibitem[{\citenamefont{Robison et~al.}(1998)\citenamefont{Robison, McGuire,
  and Church}}]{church}
\bibinfo{author}{\bibfnamefont{K.}~\bibnamefont{Robison}},
  \bibinfo{author}{\bibfnamefont{A.~M.} \bibnamefont{McGuire}},
  \bibnamefont{and} \bibinfo{author}{\bibfnamefont{G.~M.}
  \bibnamefont{Church}}, \bibinfo{journal}{Journal of Molecular Biology}
  \textbf{\bibinfo{volume}{284}}, \bibinfo{pages}{241} (\bibinfo{year}{1998}),
  \bibinfo{note}{http://arep.med.harvard.edu/dpinteract}.

\bibitem[{\citenamefont{Salgado
  et~al.}(2000{\natexlab{a}})\citenamefont{Salgado, Santos-Zavaleta,
  Gama-Castro, Millan-Zarate, Blattner, and Collado-Vides}}]{regulonDB}
\bibinfo{author}{\bibfnamefont{H.}~\bibnamefont{Salgado}},
  \bibinfo{author}{\bibfnamefont{A.}~\bibnamefont{Santos-Zavaleta}},
  \bibinfo{author}{\bibfnamefont{S.}~\bibnamefont{Gama-Castro}},
  \bibinfo{author}{\bibfnamefont{D.}~\bibnamefont{Millan-Zarate}},
  \bibinfo{author}{\bibfnamefont{F.}~\bibnamefont{Blattner}}, \bibnamefont{and}
  \bibinfo{author}{\bibfnamefont{J.}~\bibnamefont{Collado-Vides}},
  \bibinfo{journal}{Nucl. Acids Res.} \textbf{\bibinfo{volume}{28}},
  \bibinfo{pages}{65} (\bibinfo{year}{2000}{\natexlab{a}}).

\bibitem[{\citenamefont{Salgado
  et~al.}(2000{\natexlab{b}})\citenamefont{Salgado, Moreno-Hagelsieb, Smith,
  and Collado-Vides}}]{salgado}
\bibinfo{author}{\bibfnamefont{H.}~\bibnamefont{Salgado}},
  \bibinfo{author}{\bibfnamefont{G.}~\bibnamefont{Moreno-Hagelsieb}},
  \bibinfo{author}{\bibfnamefont{T.}~\bibnamefont{Smith}}, \bibnamefont{and}
  \bibinfo{author}{\bibfnamefont{J.}~\bibnamefont{Collado-Vides}},
  \bibinfo{journal}{Proc Natl. Acad. Sci USA} \textbf{\bibinfo{volume}{97}},
  \bibinfo{pages}{6652} (\bibinfo{year}{2000}{\natexlab{b}}).

\bibitem[{\citenamefont{Stormo and Hartzell}(1989)}]{consensus}
\bibinfo{author}{\bibfnamefont{G.~D.} \bibnamefont{Stormo}} \bibnamefont{and}
  \bibinfo{author}{\bibfnamefont{G.~W.} \bibnamefont{Hartzell}},
  \bibinfo{journal}{Proc. Natl. Acad. Sci. U.S.A.}
  \textbf{\bibinfo{volume}{86}}, \bibinfo{pages}{1183} (\bibinfo{year}{1989}).

\bibitem[{\citenamefont{Lawrence et~al.}(1993)\citenamefont{Lawrence, Altschul,
  Boguski, Liu, Neuwald, and Wootton}}]{Gibbssampler}
\bibinfo{author}{\bibfnamefont{C.~E.} \bibnamefont{Lawrence}},
  \bibinfo{author}{\bibfnamefont{S.~F.} \bibnamefont{Altschul}},
  \bibinfo{author}{\bibfnamefont{M.~S.} \bibnamefont{Boguski}},
  \bibinfo{author}{\bibfnamefont{J.~S.} \bibnamefont{Liu}},
  \bibinfo{author}{\bibfnamefont{A.~F.} \bibnamefont{Neuwald}},
  \bibnamefont{and} \bibinfo{author}{\bibfnamefont{J.~C.}
  \bibnamefont{Wootton}}, \bibinfo{journal}{Science}
  \textbf{\bibinfo{volume}{262}}, \bibinfo{pages}{208} (\bibinfo{year}{1993}).

\bibitem[{\citenamefont{Bailey and Elkan}(1994)}]{meme}
\bibinfo{author}{\bibfnamefont{T.}~\bibnamefont{Bailey}} \bibnamefont{and}
  \bibinfo{author}{\bibfnamefont{C.}~\bibnamefont{Elkan}},
  \bibinfo{journal}{Proceedings ISMB'94} \textbf{\bibinfo{volume}{2}},
  \bibinfo{pages}{28} (\bibinfo{year}{1994}).

\bibitem[{\citenamefont{Bussemaker et~al.}(2000)\citenamefont{Bussemaker, Li,
  and Siggia}}]{bussemaker}
\bibinfo{author}{\bibfnamefont{H.~J.} \bibnamefont{Bussemaker}},
  \bibinfo{author}{\bibfnamefont{H.}~\bibnamefont{Li}}, \bibnamefont{and}
  \bibinfo{author}{\bibfnamefont{E.~D.} \bibnamefont{Siggia}},
  \bibinfo{journal}{Proc. Natl. Acad. Sci. U.S.A.}
  \textbf{\bibinfo{volume}{97}}, \bibinfo{pages}{10096} (\bibinfo{year}{2000}).

\bibitem[{\citenamefont{Hardison et~al.}(1997)\citenamefont{Hardison, Oeltjen,
  and Miller}}]{hardison}
\bibinfo{author}{\bibfnamefont{R.}~\bibnamefont{Hardison}},
  \bibinfo{author}{\bibfnamefont{J.}~\bibnamefont{Oeltjen}}, \bibnamefont{and}
  \bibinfo{author}{\bibfnamefont{W.}~\bibnamefont{Miller}},
  \bibinfo{journal}{Genome Res.} \textbf{\bibinfo{volume}{10}},
  \bibinfo{pages}{959} (\bibinfo{year}{1997}).

\bibitem[{\citenamefont{Gelfand et~al.}(2000)\citenamefont{Gelfand, Koonin, and
  Mironov}}]{gelfand}
\bibinfo{author}{\bibfnamefont{M.}~\bibnamefont{Gelfand}},
  \bibinfo{author}{\bibfnamefont{E.}~\bibnamefont{Koonin}}, \bibnamefont{and}
  \bibinfo{author}{\bibfnamefont{A.}~\bibnamefont{Mironov}},
  \bibinfo{journal}{Nucleic Acids Res.} \textbf{\bibinfo{volume}{28}},
  \bibinfo{pages}{695} (\bibinfo{year}{2000}).

\bibitem[{\citenamefont{McGuire et~al.}(2000)\citenamefont{McGuire, Hughes, and
  Church}}]{mcguire}
\bibinfo{author}{\bibfnamefont{A.~M.} \bibnamefont{McGuire}},
  \bibinfo{author}{\bibfnamefont{J.~D.} \bibnamefont{Hughes}},
  \bibnamefont{and} \bibinfo{author}{\bibfnamefont{G.~M.}
  \bibnamefont{Church}}, \bibinfo{journal}{Genome Res.}
  \textbf{\bibinfo{volume}{10}}, \bibinfo{pages}{744} (\bibinfo{year}{2000}).

\bibitem[{\citenamefont{Berg and von Hippel}(1987)}]{von_hippel}
\bibinfo{author}{\bibfnamefont{O.~G.} \bibnamefont{Berg}} \bibnamefont{and}
  \bibinfo{author}{\bibfnamefont{P.~H.} \bibnamefont{von Hippel}},
  \bibinfo{journal}{J. Mol. Biol.} \textbf{\bibinfo{volume}{193}},
  \bibinfo{pages}{723} (\bibinfo{year}{1987}).

\bibitem[{\citenamefont{de~Bruijn}(1958)}]{DeBruijnbook}
\bibinfo{author}{\bibfnamefont{N.~G.} \bibnamefont{de~Bruijn}},
  \emph{\bibinfo{title}{Asymptotic Methods in Analysis}}
  (\bibinfo{publisher}{Dover Inc.}, \bibinfo{year}{1958}).

\bibitem[{\citenamefont{Metropolis et~al.}(1953)\citenamefont{Metropolis,
  Rosenbluth, Rosenbluth, Teller, and Teller}}]{Metropolis53}
\bibinfo{author}{\bibfnamefont{N.}~\bibnamefont{Metropolis}},
  \bibinfo{author}{\bibfnamefont{A.~W.} \bibnamefont{Rosenbluth}},
  \bibinfo{author}{\bibfnamefont{M.~N.} \bibnamefont{Rosenbluth}},
  \bibinfo{author}{\bibfnamefont{A.~H.} \bibnamefont{Teller}},
  \bibnamefont{and} \bibinfo{author}{\bibfnamefont{E.}~\bibnamefont{Teller}},
  \bibinfo{journal}{J. Chem. Phys} \textbf{\bibinfo{volume}{21}},
  \bibinfo{pages}{1087} (\bibinfo{year}{1953}).

\bibitem[{website()}]{website}
website, \bibinfo{note}{all clusters can be found at
  http://www.physics.rockefeller.edu/$\sim$erik/website\_explanation.html}.

\bibitem[{\citenamefont{Begley et~al.}(1999)\citenamefont{Begley, Downs,
  Ealick, McLafferty, and van Loon~et. al.}}]{begley}
\bibinfo{author}{\bibfnamefont{T.}~\bibnamefont{Begley}},
  \bibinfo{author}{\bibfnamefont{D.}~\bibnamefont{Downs}},
  \bibinfo{author}{\bibfnamefont{S.}~\bibnamefont{Ealick}},
  \bibinfo{author}{\bibfnamefont{F.}~\bibnamefont{McLafferty}},
  \bibnamefont{and} \bibinfo{author}{\bibfnamefont{A.}~\bibnamefont{van
  Loon~et. al.}}, \bibinfo{journal}{Arch. Microbiol.}
  \textbf{\bibinfo{volume}{171}}, \bibinfo{pages}{293} (\bibinfo{year}{1999}).

\bibitem[{\citenamefont{Miranda-Rios et~al.}(2001)\citenamefont{Miranda-Rios,
  Navarro, and Sober{\'o}n}}]{MirandaEtAl2001}
\bibinfo{author}{\bibfnamefont{J.}~\bibnamefont{Miranda-Rios}},
  \bibinfo{author}{\bibfnamefont{M.}~\bibnamefont{Navarro}}, \bibnamefont{and}
  \bibinfo{author}{\bibfnamefont{M.}~\bibnamefont{Sober{\'o}n}},
  \bibinfo{journal}{Proceedings of the National Academy of Sciences}
  \textbf{\bibinfo{volume}{98}}, \bibinfo{pages}{9736} (\bibinfo{year}{2001}).

\bibitem[{\citenamefont{Bausch et~al.}(1998)\citenamefont{Bausch, Peekhaus,
  Utz, Blais, Murray, Lowary, and Conway}}]{BauschEtAl1998}
\bibinfo{author}{\bibfnamefont{C.}~\bibnamefont{Bausch}},
  \bibinfo{author}{\bibfnamefont{N.}~\bibnamefont{Peekhaus}},
  \bibinfo{author}{\bibfnamefont{C.}~\bibnamefont{Utz}},
  \bibinfo{author}{\bibfnamefont{T.}~\bibnamefont{Blais}},
  \bibinfo{author}{\bibfnamefont{E.}~\bibnamefont{Murray}},
  \bibinfo{author}{\bibfnamefont{T.}~\bibnamefont{Lowary}}, \bibnamefont{and}
  \bibinfo{author}{\bibfnamefont{T.}~\bibnamefont{Conway}},
  \bibinfo{journal}{Journal of Bacteriology} \textbf{\bibinfo{volume}{180}},
  \bibinfo{pages}{3704} (\bibinfo{year}{1998}).

\bibitem[{\citenamefont{Peekhaus and Conway}(1998{\natexlab{a}})}]{peekhaus1}
\bibinfo{author}{\bibfnamefont{N.}~\bibnamefont{Peekhaus}} \bibnamefont{and}
  \bibinfo{author}{\bibfnamefont{T.}~\bibnamefont{Conway}},
  \bibinfo{journal}{J. Bacteriol.} \textbf{\bibinfo{volume}{180}},
  \bibinfo{pages}{3495} (\bibinfo{year}{1998}{\natexlab{a}}).

\bibitem[{\citenamefont{Peekhaus and Conway}(1998{\natexlab{b}})}]{peekhaus2}
\bibinfo{author}{\bibfnamefont{N.}~\bibnamefont{Peekhaus}} \bibnamefont{and}
  \bibinfo{author}{\bibfnamefont{T.}~\bibnamefont{Conway}},
  \bibinfo{journal}{J. Bacteriol.} \textbf{\bibinfo{volume}{180}},
  \bibinfo{pages}{1777} (\bibinfo{year}{1998}{\natexlab{b}}).

\bibitem[{\citenamefont{Neidhardt}(1996)}]{neidhardt}
\bibinfo{editor}{\bibfnamefont{F.~C.} \bibnamefont{Neidhardt}}, ed.,
  \emph{\bibinfo{title}{Escherichia coli and Salmonella Typhimurium. Cellular
  and molecular biology}} (\bibinfo{publisher}{ASM press},
  \bibinfo{address}{Washington D.C.}, \bibinfo{year}{1996}).

\bibitem[{\citenamefont{Jacobson and Fuchs}(1998)}]{Jacobson&Fuchs1998}
\bibinfo{author}{\bibfnamefont{B.~A.} \bibnamefont{Jacobson}} \bibnamefont{and}
  \bibinfo{author}{\bibfnamefont{J.~A.} \bibnamefont{Fuchs}},
  \bibinfo{journal}{Molecular Microbiology} \textbf{\bibinfo{volume}{28}},
  \bibinfo{pages}{1315} (\bibinfo{year}{1998}).

\bibitem[{Ecocyc()}]{EcoCyc}
Ecocyc, \emph{\bibinfo{title}{The ecocyc database}},
  \bibinfo{note}{http://ecocyc.pangeasystems.com/}.

\bibitem[{\citenamefont{Jennings and Beacham}(1990)}]{jennings}
\bibinfo{author}{\bibfnamefont{M.}~\bibnamefont{Jennings}} \bibnamefont{and}
  \bibinfo{author}{\bibfnamefont{I.}~\bibnamefont{Beacham}},
  \bibinfo{journal}{J. Bacteriol.} \textbf{\bibinfo{volume}{172}},
  \bibinfo{pages}{1491} (\bibinfo{year}{1990}).

\bibitem[{\citenamefont{Cheek and Broderick}(2001)}]{cheek}
\bibinfo{author}{\bibfnamefont{J.}~\bibnamefont{Cheek}} \bibnamefont{and}
  \bibinfo{author}{\bibfnamefont{J.}~\bibnamefont{Broderick}},
  \bibinfo{journal}{J. Biol. Inorg. Chem.} \textbf{\bibinfo{volume}{6}},
  \bibinfo{pages}{209} (\bibinfo{year}{2001}).

\bibitem[{\citenamefont{van Nimwegen et~al.}(1999)\citenamefont{van Nimwegen,
  Crutchfield, and Huynen}}]{NimwEtAl99}
\bibinfo{author}{\bibfnamefont{E.}~\bibnamefont{van Nimwegen}},
  \bibinfo{author}{\bibfnamefont{J.~P.} \bibnamefont{Crutchfield}},
  \bibnamefont{and} \bibinfo{author}{\bibfnamefont{M.}~\bibnamefont{Huynen}},
  \bibinfo{journal}{Proc. Natl. Acad. Sci. USA} \textbf{\bibinfo{volume}{96}},
  \bibinfo{pages}{9716} (\bibinfo{year}{1999}).

\bibitem[{\citenamefont{Sengupta et~al.}(2002)\citenamefont{Sengupta,
  Djordjevic, and Shraiman}}]{Borispaper}
\bibinfo{author}{\bibfnamefont{A.~M.} \bibnamefont{Sengupta}},
  \bibinfo{author}{\bibfnamefont{M.}~\bibnamefont{Djordjevic}},
  \bibnamefont{and} \bibinfo{author}{\bibfnamefont{B.~I.}
  \bibnamefont{Shraiman}}, \bibinfo{journal}{Proc. Natl. Acad. Sci. USA}
  \textbf{\bibinfo{volume}{99}}, \bibinfo{pages}{2072} (\bibinfo{year}{2002}).

\bibitem[{\citenamefont{Jaynes}(1983)}]{Jaynescollection}
\bibinfo{author}{\bibfnamefont{E.~T.} \bibnamefont{Jaynes}},
  \emph{\bibinfo{title}{Papers on probability, statistics, and statistical
  physics}}, vol. \bibinfo{volume}{158} of \emph{\bibinfo{series}{Synthese
  library}} (\bibinfo{publisher}{D. Reidel publishing company},
  \bibinfo{address}{Dordrecht Holland}, \bibinfo{year}{1983}),
  \bibinfo{note}{r.D. Rosenkrantz, editor}.

\end{thebibliography}

\end{document}